\documentclass[12pt,a4paper]{article}
\usepackage[T1]{fontenc}
\usepackage{fancyhdr}
\usepackage{amsmath,amsthm,amsfonts,amssymb}
\allowdisplaybreaks
\usepackage{epic,eepic}
\usepackage{graphicx}
\usepackage{authblk}
\usepackage{hyperref}
\usepackage{fullpage}
\usepackage[active]{srcltx}
\usepackage{enumitem}
\usepackage{booktabs}

\usepackage{framed,color}

\usepackage{float}%
\floatstyle{plaintop}%
\restylefloat{table}

\usepackage[font={small,it}]{caption}

\usepackage[table]{xcolor}

\usepackage[toc,page]{appendix}

\usepackage[style=authoryear,backend=bibtex,maxcitenames=2,maxbibnames=99]{biblatex}
\bibliography{SBTM}

\usepackage{chngcntr}
\usepackage{dsfont}

\usepackage{algpseudocode}
\usepackage{algorithm}
\usepackage{setspace}

\usepackage{tikz}
\usetikzlibrary{bayesnet}

\makeatletter
\AtEveryBibitem{%
  \global\undef\bbx@lasthash%
  \clearfield{}}
\makeatother

\theoremstyle{plain}

\theoremstyle{definition}

\theoremstyle{remark}

\numberwithin{equation}{section}
\numberwithin{figure}{section}

\date{}
\title{\textbf{Exact integrated completed likelihood maximisation in a stochastic block transition model for dynamic networks}}
\author{Riccardo Rastelli}
\affil{\footnotesize riccardo.rastelli@wu.ac.at}
\affil{\footnotesize Institute for Statistics and Mathematics, Vienna University of Economics and Business, Austria.}

\begin{document}
\rowcolors{2}{gray!25}{white}
\counterwithout{figure}{section}
\counterwithout{figure}{subsection}
\counterwithout{equation}{section}
\counterwithout{equation}{subsection}

\maketitle
\begin{abstract}
\noindent
The latent stochastic block model is a flexible and widely used statistical model for the analysis of network data. 
Extensions of this model to a dynamic context often fail to capture the persistence of edges in contiguous network snapshots. 
The recently introduced stochastic block transition model addresses precisely this issue, by modelling the probabilities of creating a new edge and of maintaining an edge over time. 
Using a model-based clustering approach, this paper illustrates a methodology to fit stochastic block transition models under a Bayesian framework. 
The method relies on a greedy optimisation procedure to maximise the exact integrated completed likelihood. 
The computational efficiency of the algorithm used makes the methodology scalable and appropriate for the analysis of large network datasets. 
Crucially, the optimal number of latent groups is automatically selected at no additional computing cost. 
The efficacy of the method is demonstrated through applications to both artificial and real datasets.
\\

\end{abstract}

\baselineskip=20pt
\section{Introduction}\label{sec:introduction}
Research on networks has gained significant momentum in the last few decades. 
In fact, networks can be used to represent observed phenomena in a variety of research areas, including the social sciences, epidemiology, biology, technology and finance. 
Social networks, such as collaboration networks or proximity networks, are largely available, 
and they pose stimulating research challenges, since they typically require scalable and well-thought statistical methodologies.

Most frequently, network data is provided in the form of an adjacency matrix, where each entry $x_{ij}$ characterises the interaction between the nodes $i$ and $j$.
The Stochastic Block Model (SBM), as characterised by \textcite{wang1987}, is a flexible statistical model that can be used to analyse large social networks.
In the SBM, the nodes of the network are assigned to latent groups based on their connection preferences:
two nodes belonging to the same group are said stochastically equivalent, meaning that they have the same probability of connecting to any other node in the network.
This concept generalises the idea of community structure (see \textcite{fortunato2010community} and references therein), since disassortative behaviours may be represented.
The SBM framework effectively defines a clustering problem, where one has to estimate from the data both the nodes' cluster membership variables (\textit{allocations}) and the underlying number of clusters.
This may be tackled in a number of ways, including sampling based approaches \parencite{nowicki2001estimation}, or optimisation approaches relying on the Expectation-Maximisation algorithm \parencite{daudin2008mixture}.
A recent survey on SBMs can be found in \textcite{matias2014modeling}.


In recent years, a number of works have extended the \textit{static} SBM to the \textit{dynamic} framework, whereby the observed interactions are, in some way, dynamically evolving over time.
One type of extension considers the interactions as instantaneous events which may be observed in any given instant.
For example, \textcite{matias2015semiparametric} and \textcite{corneli2017multiple} model these interactions as realised events of a non-homogeneous Poisson point process, 
where the intensity parameters are determined by the cluster memberships of the corresponding nodes.

This paper belongs to a different strand of literature, where the time dimension is discretised and the observed data can be represented as a collection of adjacency matrices indexed according to their ordering in time.
Most of the works following this approach typically introduce a Markov property that creates a temporal dependency between any two contiguous network snapshots.
For example, \textcite{yang2011detecting} assume a hidden Markov model where the hidden states are the cluster membership variables of the nodes.
In their model, the temporal dependency is captured only through the evolution of the latent allocation variables over time.
By contrast, \textcite{xu2014dynamic} characterise the time dependency through a state-space model on the connection probabilities between the SBM blocks.
\textcite{matias2017statistical} consider a more general framework that includes the previous two as special cases, 
proving that the identifiability of these Markovian models may be lost if both cluster membership variables and connectivity parameters are allowed to change over time.
They also propose an estimation method that can handle networks with non-binary interactions.
\textcite{rastelli2017choosing} also focus on the same type of model, studying the computational efficiency and scalability of the inferential process.
Other relevant dynamic extensions of the SBM are introduced in \textcite{ishiguro2010dynamic} and \textcite{bartolucci2018dealing}.


In many cases, however, the observed dynamic networks tend to be particularly stable over time, or, equivalently, they exhibit a strong temporal dependency.
This may have important repercussions, since it ultimately questions whether the temporal dynamics of the models are necessary, or if the static frameworks may be just as effective.
The dynamic SBM models mentioned so far are not able to capture these additional temporal dependencies.
\textcite{xu2015stochastic} addresses exactly this issue, proposing an original model that builds upon a dynamic SBM to include a Markov property on the observed edge values.
Differently from the SBM structure, this model clusters the nodes in each time frame according to their propensity to create new edges and maintaining existing ones.
Since it directly models the transitions of the edge values, it is called the Stochastic Block Transition Model (SBTM). 
In the SBTM, the probability of observing an edge depends on whether the same edge was present or absent in the previous time frame, creating a direct dependency between any two contiguous network snapshots.
\textcite{xu2015stochastic} gives evidence that the SBTM can successfully model the creation and the duration of the interactions, hence being much more flexible than the dynamic SBM of \textcite{xu2014dynamic}.

Although in different modelling contexts, other works propose similar ideas: 
\textcite{friel2016} consider an extension of the Latent Position Model for bipartite dynamic networks, and they introduce additional parameters to explicitly capture the persistence of edges over time.
\textcite{zhang2017random}, instead, study a model similar to the SBTM, obtained through a discretisation of an underlying continuous-time process. 
They consider a framework that facilitates both the theoretical characterisation of such model and a likelihood-based inferential procedure.
Finally, \textcite{heaukulani2013dynamic} propose a type of block model where the evolution over time of the latent allocation of each node is affected by the cluster memberships of its neighbours.\\

This paper focuses on the SBTM, and it extends the work of \textcite{xu2015stochastic} in a number of ways. 
First, a new Bayesian hierarchical structure for this model is introduced, following ideas similar to those in \textcite{rastelli2017choosing}.
The generative process proposed allows for non-informative priors and, crucially, it directly captures the fact that nodes may become inactive in certain time intervals.
This feature makes the model proposed particularly suitable for the analysis of longitudinal network data, whereby some nodes are added or removed at each time frame.
Then, the modelling assumptions are exploited to analytically integrate out (\textit{collapse}) most of the model parameters, as also advocated by \textcite{nobile2007bayesian, mcdaid2013improved, come2015model}.
This collapsing leads to an exact formula for the well known Integrated Completed Likelihood (ICL), which is widely used as an optimality criterion in the statistical analysis of finite mixtures \parencite{biernacki2000assessing}.
The exact ICL value obtained is maximised with respect to the allocation variables using a scalable heuristic greedy procedure, 
which resembles the algorithms described by \textcite{come2015model,wyse2017inferring,rastelli2017choosing}.

An important advantage of the methodology proposed is that the number of latent groups can be automatically deduced from the allocation variables at any stage of the optimisation.
In fact, to the best of my knowledge, this is currently the only paper addressing the problem of model choice for the SBTM.
Another facet of this method is that, due to the optimisation context, it is unaffected by label-switching issues.
In addition, the algorithm can exploit the presence of inactive nodes, which further reduces the computational burden.

Taking advantage of the non-informative setting, the methodology is tested as a black-box tool on artificially generated networks, showing that it achieves good convergence and an overall good clustering performance.
The procedure is also compared with other available methods, showing that the introduction of the edge-persistence feature is essential to recover the true partitioning of the nodes and the correct generative mechanism.
In addition, a large longitudinal human contact dataset is used to give a demonstration of the procedure, showing that the results obtained are easy to interpret, 
and that the behaviours of the nodes can be analysed in detail.

Finally, the \texttt{R} package \texttt{GreedySBTM} accompanies this paper and it provides an implementation of the algorithm described. 
The package is publicly available on \texttt{CRAN} \parencite{rcoreteam}.

The paper is organised as follows: Sections \ref{sec:SBTM} and \ref{sec:hierarchical} illustrate the Bayesian SBTM, 
Sections \ref{sec:icl} and \ref{sec:greedy} describe the exact ICL approach and the optimisation algorithm,
and finally the methodology is applied to simulated and a real dataset in Sections \ref{sec:simulations} and \ref{sec:reality}, respectively.

\section{The Stochastic Block Transition Model}\label{sec:SBTM}
The statistical model used in this paper is a variation of that introduced by \textcite{xu2015stochastic}.
The differences between the two models are minor and do not affect the principle ideas that motivate the use of the SBTM;
nonetheless they are necessary to give integrity to the inferential procedure used in this paper.
A more detailed account of the modifications and a comparison with other statistical models for dynamic network data is provided at the end of the next section.

The observed data consist of a collection of $T$ graphs, where the edges of each of these represent interactions between the corresponding nodes at different times.
In each time frame $t=\left\{1,\dots,T \right\}$, some of the nodes may be \textit{inactive}, in which case none of their edge values are observed, or they simply do not have any interaction.
The data may be described through two binary cubes $\mathcal{X}$ and $\mathcal{Y}$ of size $N\times N\times T$, which are characterised by:
\begin{equation}
\rowcolors{1}{}{}
y_{ij}^{(t)} = \begin{cases}
                 1&\mbox{ if both nodes $i$ and $j$ are active at time frame $t$},\\
                 0&\mbox{ otherwise};
                \end{cases}
\end{equation}
\begin{equation}
\rowcolors{1}{}{}
x_{ij}^{(t)} = \begin{cases}
                 1&\mbox{ if }y_{ij}^{(t)} = 1\mbox{ \textbf{and} an edge between $i$ and $j$ appears at time $t$},\\
                 0&\mbox{ if }y_{ij}^{(t)} = 0\mbox{ \textbf{or} no edge appears between $i$ and $j$ at time $t$},
                \end{cases}
\end{equation}
for all $i$ and $j$ in $\left\{ 1,\dots,N\right\}$ and $t$ in $\left\{1,\dots,T \right\}$.
Evidently, $\mathcal{Y}$ simply serves as an activity indicator, whereas $\mathcal{X} = \left\{ \textbf{X}^{(1)},\dots, \textbf{X}^{(T)}\right\}$ corresponds to a collection of canonical adjacency matrices for the observed edge values.
The $T$ graphs are assumed to be undirected and without self-edges, hence each of the adjacency matrices is symmetric and has zeros on the diagonal.

In the SBTM, a clustering structure is hypothesised on the nodes of the $T$ observed graphs.
Each of the nodes, at each time frame, is characterised by a cluster membership variable taking values in the discrete set $\left\{ 0, 1,\dots, K\right\}$.
The notation $\mathcal{Z} = \left\{ z_i^{(t)}:\ i=1,\dots,N,\ t=1,\dots,T\right\}$ is used to denote all such allocations.
Also, the equivalent notation $z_{ig}^{(t)} = \mathds{1}(\{ z_{i}^{(t)} = g \})$ may be used in some equations ($\mathds{1}$ denotes the indicator function).
Note that the vector $\textbf{z}^{(t)} = \left\{z_1^{(t)},\dots,z_N^{(t)}\right\}$ denotes a partition of $\left\{1,\dots,N\right\}$, for every $t$.
Finally, the label zero is reserved for inactive nodes, i.e. $z_i^{(t)} = 0$ iff $i$ is inactive at time $t$: 
since the inactive nodes are known, there is no interest in inferring these allocations and hence they are kept fixed throughout.

The probability that the observed edge indexed by $\left( i,j,t \right)$ takes value $1$ is defined as:
\begin{equation}\label{eq:pijt}
\rowcolors{1}{}{}
\begin{split}
\rho_{ij}^{(t)} &= \mathbb{P}\left( x_{ij}^{(t)} = 1 \middle\vert y_{ij}^{(t)} = 1, y_{ij}^{(t-1)}, x_{ij}^{(t-1)}, z_{i}^{(t)} = g, z_{j}^{(t)} = h, \theta_{gh}, P_{gh}, Q_{gh} \right) \\
&= 1 - \mathbb{P}\left( x_{ij}^{(t)} = 0 \middle\vert y_{ij}^{(t)} = 1, y_{ij}^{(t-1)}, x_{ij}^{(t-1)}, z_{i}^{(t)} = g, z_{j}^{(t)} = h, \theta_{gh}, P_{gh}, Q_{gh} \right) \\
 &= \begin{cases}
     \theta_{gh} & \mbox{ if } y_{ij}^{(t-1)} = 0 \\
     P_{gh} & \mbox{ if } y_{ij}^{(t-1)} = 1 \mbox{ and } x_{ij}^{(t-1)} = 0 \\
     1-Q_{gh} & \mbox{ if } y_{ij}^{(t-1)} = 1 \mbox{ and } x_{ij}^{(t-1)} = 1.
    \end{cases}
\end{split}
\end{equation}
Note that if $t = 1$ then $y_{ij}^{(t-1)} = 0$ for all $i$ and $j$. 
The probability of an edge $\rho_{ij}^{(t)}$ is defined by \eqref{eq:pijt} only if $y_{ij}^{(t)} = 1$, in fact, only observed edges may contribute to the likelihood value.
Equation \eqref{eq:pijt} essentially characterises the alternation of three regimes: 
\begin{itemize}
 \item A SBM-type of connection probability $\theta_{gh}$ is selected whenever there is no information regarding the previous value of the edge considered.
 \item A SBTM probability $P_{gh}$ is used when it is known that the edge considered had value zero in the previous time frame. 
 The value $P_{gh}$ corresponds to the probability of creating a new edge.
 \item A SBTM probability $Q_{gh}$ is used when it is known that the edge considered had value one in the previous time frame. 
 The probability of confirming the edge is $1-Q_{gh}$, hence $Q_{gh}$ may be interpreted as the probability of destroying an existing edge.
\end{itemize}
These parameters $\left\{\theta_{gh}\right\}_{g,h}$, $\left\{P_{gh}\right\}_{g,h}$ and $\left\{Q_{gh}\right\}_{g,h}$ form the matrices $\boldsymbol{\Theta}$, $\textbf{P}$ and $\textbf{Q}$ respectively, 
which contain the edge probabilities for nodes belonging to any two given groups.

The full likelihood of the model can be written as:
\begin{equation}\label{eq:likelihood1}
\begin{split}
 \mathcal{L}_{\mathcal{X},\mathcal{Y}}\left( \mathcal{Z},\boldsymbol{\Theta},\textbf{P},\textbf{Q} \right) 
 &= \prod_{t=1}^{T}\prod_{i<j} \left\{\left[ p_{ij}^{(t)} \right]^{x_{ij}^{(t)}}  \left[ 1-p_{ij}^{(t)} \right]^{1-x_{ij}^{(t)}}\right\}^{y_{ij}^{(t)}}\\
\end{split}
\end{equation}
which is simply a product of contributions given by Bernoulli variables. 
Hereafter, the product $\prod_{i<j}$ stands for $\prod_{i=1}^{N-1}\prod_{j=i+1}^{N}$, for brevity.
The likelihood function may be reformulated in a more convenient way, taking advantage of the block structure and hence grouping up the terms in \eqref{eq:likelihood1}.
In order to do this, the following quantities are needed, for all $g$ and $h$ in $\left\{1,\dots,K\right\}$:
\begin{equation}\label{eq:eta}
\eta_{gh}  = \sum_{i<j} y_{ij}^{(1)}x_{ij}^{(1)}\lambda_{ij1gh} + \sum_{t>1}\sum_{i<j}y_{ij}^{(t)}\left( 1-y_{ij}^{(t-1)} \right)x_{ij}^{(t)} \lambda_{ijtgh};
\end{equation}
\begin{equation}\label{eq:zeta}
\zeta_{gh}  = \sum_{i<j} y_{ij}^{(1)}\left(1-x_{ij}^{(1)}\right)\lambda_{ijtgh} + \sum_{t>1}\sum_{i<j}y_{ij}^{(t)}\left( 1-y_{ij}^{(t-1)} \right)\left(1-x_{ij}^{(1)}\right)\lambda_{ijtgh};
\end{equation}
\begin{equation}\label{eq:u}
 U_{gh}^{uv} = \sum_{t>1}\sum_{i<j} y_{ij}^{(t)}y_{ij}^{(t-1)}\left[1-\left( u-x_{ij}^{(t-1)} \right)^2\right]\left[1-\left( v-x_{ij}^{(t)} \right)^2\right]\lambda_{ijtgh};
\end{equation}
\begin{equation}
\rowcolors{1}{}{}
\begin{split}
 \lambda_{ijtgh} &= z_{ig}^{(t)}z_{jh}^{(t)} + z_{ih}^{(t)}z_{jg}^{(t)} - z_{ig}^{(t)}z_{jg}^{(t)}z_{ih}^{(t)}z_{jh}^{(t)} \\
 &= \begin{cases}
     1&\mbox{ if $\left( i,j,t \right)$ refers to an edge between groups $g$ and $h$};\\
     0&\mbox{ otherwise}.\\
    \end{cases}
\end{split}
\end{equation}
The values $u$ and $v$ are in $\left\{0,\ 1\right\}$. 
Also note that for binary values $c_1$ and $c_2$:
\begin{equation}
\rowcolors{1}{}{}
 1-\left( c_1-c_2 \right)^2 = \begin{cases}
  1&\mbox{ if }c_1=c_2;\\
  0&\mbox{ otherwise. }
 \end{cases}
\end{equation}
The quantities introduced in \eqref{eq:eta}, \eqref{eq:zeta} and \eqref{eq:u} are crucial summaries of the data. 
They can be interpreted as the number of successes in creating a SBM-edge ($\eta_{gh}$), in creating a new edge ($U_{gh}^{01}$), and destroying an existing edge ($U_{gh}^{10}$), between a node in group $g$ and one in group $h$. 
Similarly, $\zeta_{gh}$, $U_{gh}^{00}$ and $U_{gh}^{11}$ correspond to the number of failures for the same events, respectively.
Using these new quantities, the likelihood function factorises as follows:
\begin{equation}
 \mathcal{L}_{\mathcal{X},\mathcal{Y}}\left( \mathcal{Z},\boldsymbol{\Theta},\textbf{P},\textbf{Q} \right) 
 = \prod_{g=1}^{K}\prod_{h=g}^{K} \theta_{gh}^{\eta_{gh}}\left( 1-\theta_{gh} \right)^{\zeta_{gh}}  P_{gh}^{U_{gh}^{01}}\left( 1-P_{gh} \right)^{U_{gh}^{00}}  Q_{gh}^{U_{gh}^{11}}\left( 1-Q_{gh} \right)^{U_{gh}^{10}}.
\end{equation}
This likelihood formulation is similar to those proposed by \textcite{xu2015stochastic} and \textcite{zhang2017random}.
Exactly as in SBMs, the presence of blocks simplifies the model structure, and can be exploited to design efficient inferential procedures.

\section{Bayesian hierarchical structure}\label{sec:hierarchical}
This section introduces a Bayesian hierarchical structure for the SBTM, hence proposing a generative mechanism for the observed data.
The prior distributions described here are all conjugate, and, as a special case, they permit a non-informative framework which may be used to nullify the subjective contribution induced by the user.

As concerns the allocations, these are assumed to evolve according to $N$ independent Markov chains on the states $\left\{0,\dots,K\right\}$.
The processes share the same transition probability matrix $\boldsymbol{\Pi}$, which is hence characterised by:
\begin{equation}
 \pi_{gh} = \mathbb{P}\left( z_i^{(t)} = h \middle\vert z_i^{(t-1)} = g \right),
\end{equation}
for all $i=1,\dots,N$ and $t=2,\dots,T$.
The initial states are assumed to be drawn from a categorical distribution with probabilities $\alpha_0,\dots,\alpha_K$ proportional to the aggregated group sizes:
\begin{equation}
 \alpha_g \propto \sum_{t=2}^{T} \sum_{i=1}^{N} z_{ig}^{(t)}.
\end{equation}
These group proportions approximate the probabilities of the stationary distribution of the Markov chain.
Hence, the prior probability of a set of allocations $\mathcal{Z}$ may be written as follows:
\begin{equation}
 \begin{split}
  \mathbb{P}\left( \mathcal{Z}\middle\vert\boldsymbol{\Pi} \right) &= \mathbb{P}\left( \textbf{z}^{(1)}\middle\vert\boldsymbol{\alpha} \right)\prod_{t=2}^{T} \mathbb{P}\left( \textbf{z}^{(t)}\middle\vert \textbf{z}^{(t-1)}, \boldsymbol{\Pi} \right) \\
  &= \prod_{g=0}^{K} \left[\alpha_g\right]^{N_g^{(1)}}  \prod_{g=0}^{K}\prod_{h=0}^{K} \left[\pi_{gh}\right]^{R_{gh}},
 \end{split}
\end{equation}
where $N_g^{(t)} = \sum_{i=1}^{N} z_{ig}^{(t)}$ and $R_{gh} = \sum_{t=2}^{T}\sum_{i=1}^{N} z_{ig}^{(t-1)}z_{ih}^{(t)}$, for all $t$, $i$ and $g$.
Note that, while inactive nodes do not give any contribution to the likelihood in \eqref{eq:likelihood1}, their group membership affects the prior distribution at all times. 
In fact, the transition probability matrix $\boldsymbol{\Pi}$ has size $(K+1)\times(K+1)$, and it includes the probabilities for a node to be activated or inactivated.
The rows of the transition probability matrix $\boldsymbol{\Pi}$ are assumed to be independent realisations of Dirichlet random vectors:
\begin{equation}
 \left( \pi_{g0},\dots,\pi_{gK} \right) \sim Dir\left( \delta_{g0},\dots,\delta_{gK} \right),
\end{equation}
with $\delta_{gh}$ being a user-defined hyperparameter, for all $g$s and $h$s.

As concerns the likelihood parameters, the entries of the matrices $\boldsymbol{\Theta}$, $\textbf{P}$ and $\textbf{Q}$ all correspond to the success probabilities of Bernoulli random variables:
for this reason, independent Beta priors are adopted:
\begin{equation}
 \begin{split}
  \theta_{gh} &\sim Beta(\eta^0_{gh},\zeta^0_{gh}); \\
  P_{gh} &\sim Beta(a^\textbf{P}_{gh},b^\textbf{P}_{gh}); \\
  Q_{gh} &\sim Beta(a^\textbf{Q}_{gh},b^\textbf{Q}_{gh}).
 \end{split}
\end{equation}

The complete set of hyperparameters is $\boldsymbol{\phi} = \left\{ \delta_{gh}, \eta_{gh}^0, \zeta_{gh}^0, a^\textbf{P}_{gh},b^\textbf{P}_{gh}, a^\textbf{Q}_{gh},b^\textbf{Q}_{gh}\right\}_{g,h}$. 
These values should be set so that the corresponding prior distributions describe the prior knowledge available on the model parameters.
In this paper, non-informative Jeffreys' priors are assumed throughout on all model parameters: this is achieved by setting all the components of $\boldsymbol{\phi}$ to $0.5$. 

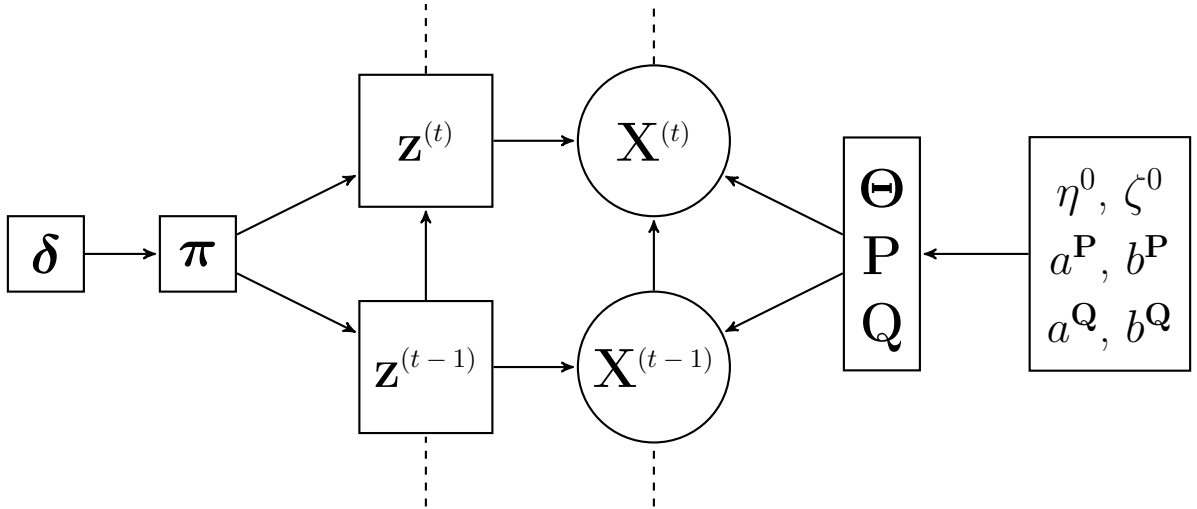
\begin{figure}[htb]
 \centering
\begin{tikzpicture}[->,>=stealth',shorten >=1pt,auto,node distance=3cm,thick,main node/.style={circle,draw,font=\sffamily\Large\bfseries}]
  \node[obs, shape=rectangle, xshift=1.7cm, fill=white, minimum size=1cm]  (delta) at (1,1.5) {\LARGE{$\boldsymbol{\delta}$}};
  \node[obs, shape=rectangle, xshift=1.7cm, fill=white, minimum size=1cm, align=center]  (phi) at (15,1.5) {\\ \\ \Large{ $\eta^0$,$\ \zeta^0\ $}\\ \\ \Large{ $a^{\textbf{P}}$,$\ b^{\textbf{P}}\ $}\\ \\ \Large{ $a^{\textbf{Q}}$,$\ b^{\textbf{Q}}\ $}\\ };
  \node[obs, shape=rectangle, xshift=1.7cm, fill=white, minimum size=1cm]  (pi) at (3,1.5) {\LARGE{$\boldsymbol{\pi}$}};
  \node[obs, shape=rectangle, xshift=1.7cm, fill=white, minimum size=1cm, align=center]  (theta) at (12,1.5) {\\ \\ \LARGE{$\boldsymbol{\Theta}$}\\ \\ \LARGE{$\textbf{P}$}\\ \\ \LARGE{$\textbf{Q}$}\\ };
  \node[obs, shape=rectangle, xshift=1.7cm, fill=white, minimum size=1.75cm]  (Ztm1) at (6,0) {\LARGE{$\textbf{z}$}\textsuperscript{\footnotesize $(t-1)$}};
  \node[obs, shape=rectangle, xshift=1.7cm, fill=white, minimum size=1.75cm]  (Zt) at (6,3) {\LARGE{$\textbf{z}$}\textsuperscript{\footnotesize $(t)$}};
  \node[obs, shape=circle, xshift=1.7cm, fill=white, minimum size=2cm]  (Xtm1) at (9,0) {\LARGE{$\textbf{X}$}\textsuperscript{\footnotesize $(t-1)$}};
  \node[obs, shape=circle, xshift=1.7cm, fill=white, minimum size=2cm]  (Xt) at (9,3) {\LARGE{$\textbf{X}$}\textsuperscript{\footnotesize $(t)$}};
  \node[] (invtopz) at (7.7,5) {};
  \node[] (invbotz) at (7.7,-2) {};
  \node[] (invtopx) at (10.7,5) {};
  \node[] (invbotx) at (10.7,-2) {};
  \draw (delta) -- (pi) node [midway, fill=white, scale=0.5, above=-0.35] {};
  \draw (phi) -- (theta) node [midway, fill=white, scale=0.5, above=-0.35] {};
  \draw (theta) -- (Xtm1) node [midway, fill=white, scale=0.5, above=-0.35] {};
  \draw (theta) -- (Xt) node [midway, fill=white, scale=0.5, above=-0.35] {};
  \draw (pi) -- (Ztm1) node [midway, fill=white, scale=0.5, above=-0.35] {};
  \draw (pi) -- (Zt) node [midway, fill=white, scale=0.5, above=-0.35] {};
  \draw (Ztm1) -- (Zt) node [midway, fill=white, scale=0, above=-0.35] {};
  \draw (Ztm1) -- (Xtm1) node [midway, fill=white, scale=0, above=-0.35] {};
  \draw (Zt) -- (Xt) node [midway, fill=white, scale=0, above=-0.35] {};
  \draw (Xtm1) -- (Xt) node [midway, fill=white, scale=0, above=-0.35] {};
  \draw [-,dashed] (Zt) to (invtopz);
  \draw [-,dashed] (Xt) to (invtopx);
  \draw [-,dashed] (invbotz) to (Ztm1);
  \draw [-,dashed] (invbotx) to (Xtm1);
\end{tikzpicture}
\caption{Graphical model for the SBTM described.}
\label{fig:graphical_model}
\end{figure}

A graphical representation of the dependencies in the model is shown in Figure \ref{fig:graphical_model}.

The Bayesian hierarchical structure introduced in this section extends the model proposed by \textcite{xu2015stochastic}.
In particular, here the allocations are allowed to change according to a Markov process, following the approaches of \textcite{matias2017statistical} and \textcite{rastelli2017choosing}.
This specification creates an additional temporal dependency, and it ultimately permits an assessment of the stability of the network.
The same feature also distinguishes this model from that of \textcite{zhang2017random}, where the allocations do not change over time.

Another important difference with the work of \textcite{xu2015stochastic} is the absence of the scaling factors, which simplifies the model and makes it more tractable.
However, as a consequence, the marginalised network snapshots are no longer guaranteed to follow a SBM structure.

\section{Exact Integrated Completed Likelihood}\label{sec:icl}
The Integrated Completed Likelihood (ICL), first introduced in \textcite{biernacki2000assessing}, is a model-based clustering criterion used to estimate the number of clusters in finite mixture models. 
In the dynamic network context addressed in this paper, the exact ICL corresponds to the following value:
\begin{equation}\label{eq:icl1}
 \mathcal{ICL}_{ex} = \mathbb{P}\left( \mathcal{X},\mathcal{Y}, \mathcal{Z} \middle\vert \boldsymbol{\phi}, K \right).
\end{equation}
Since the data $\left(\mathcal{X},\mathcal{Y}\right)$ is fixed, the $\mathcal{ICL}_{ex}$ index is also equivalent to the marginal posterior for the allocations:
\begin{equation}\label{eq:icl2}
 \mathcal{ICL}_{ex} \propto \mathbb{P}\left( \mathcal{Z} \middle\vert \mathcal{X},\mathcal{Y}, \boldsymbol{\phi}, K \right).
\end{equation}
In other words, the $\mathcal{ICL}_{ex}$ value can be obtained by analytically integrating out all of the model parameters from the full posterior distribution 
$\pi\left( \mathcal{Z},\boldsymbol{\Theta},\textbf{P},\textbf{Q},\boldsymbol{\Pi} \middle\vert \mathcal{X}, \mathcal{Y}, \boldsymbol{\phi}, K \right)$.
In fact, thanks to the conjugacy of the prior distributions, such integration is analytically possible, and the exact ICL results as follows:
\begin{equation}\label{eq:icl3}
\begin{split}
 \mathcal{ICL}_{ex} &\propto \prod_{g=0}^{K} \left\{ \left[ \alpha_g \right]^{N_{g}^{1}} \cdot \frac{\Gamma\left( \sum_{h=0}^{K}\delta_{gh} \right)}{\Gamma\left( \sum_{h=0}^{K}\delta_{gh}+\sum_{h=0}^{K}R_{gh} \right)}
 \prod_{h=0}^{K} \frac{\Gamma\left( \delta_{gh}+R_{gh} \right)}{\Gamma\left( \delta_{gh}\right)}\right\} \\
 &\cdot\prod_{g=1}^{K} \prod_{h=g}^{K} \left\{ \frac{\Gamma\left( \eta_{gh}^0+\zeta_{gh}^0 \right)}{\Gamma\left( \eta_{gh}^0 \right)\Gamma\left( \zeta_{gh}^0 \right)} 
 \cdot \frac{\Gamma\left( \eta_{gh}^0+\eta_{gh} \right)\Gamma\left( \zeta_{gh}^0+\zeta_{gh} \right)}{\Gamma\left( \eta_{gh}^0+\eta_{gh}+\zeta_{gh}^0+\zeta_{gh} \right)} \right\} \\
 &\cdot\prod_{g=1}^{K} \prod_{h=g}^{K} \left\{ \frac{\Gamma\left( a_{gh}^{\textbf{P}}+b_{gh}^{\textbf{P}} \right)}{\Gamma\left( a_{gh}^{\textbf{P}} \right)\Gamma\left( b_{gh}^{\textbf{P}} \right)} 
 \cdot \frac{\Gamma\left( a_{gh}^{\textbf{P}}+U_{gh}^{01} \right)\Gamma\left( b_{gh}^{\textbf{P}}+U_{gh}^{00} \right)}{\Gamma\left( a_{gh}^{\textbf{P}}+U_{gh}^{01}+b_{gh}^{\textbf{P}}+U_{gh}^{00} \right)} \right\} \\
 &\cdot\prod_{g=1}^{K} \prod_{h=g}^{K} \left\{ \frac{\Gamma\left( a_{gh}^{\textbf{Q}}+b_{gh}^{\textbf{Q}} \right)}{\Gamma\left( a_{gh}^{\textbf{Q}} \right)\Gamma\left( b_{gh}^{\textbf{Q}} \right)} 
 \cdot \frac{\Gamma\left( a_{gh}^{\textbf{Q}}+U_{gh}^{10} \right)\Gamma\left( b_{gh}^{\textbf{Q}}+U_{gh}^{11} \right)}{\Gamma\left( a_{gh}^{\textbf{Q}}+U_{gh}^{10}+b_{gh}^{\textbf{Q}}+U_{gh}^{11} \right)} \right\}.
\end{split}
\end{equation}


\section{Greedy optimisation}\label{sec:greedy}
The only unknown quantities in \eqref{eq:icl2} and \eqref{eq:icl3} are the allocations $\mathcal{Z}$. 
In fact, for a given clustering configuration $\mathcal{Z}$, the corresponding value of $K$ may be automatically deduced by counting the number of non empty groups.
Hence, an optimisation problem can be set up to find the allocations $\hat{\mathcal{Z}}$ maximising $\log\left(\mathcal{ICL}_{ex}\right)$,
by searching in the space of all possible clustering configurations.

This discrete optimisation problem is known to be NP-hard, and it can be solved exactly only through enumeration, which is impractical even for very small datasets.
However, heuristic greedy algorithms have been shown to perform well in similar types of clustering problems:
the procedure proposed here follows ideas similar to those of \textcite{karrer2011stochastic, come2015model, bertoletti2015choosing, rastelli2017choosing}.

First, a maximum number of groups allowed, denoted $K_{up}$, is fixed. 
For small datasets this may be set to $NT$, however, for larger networks, a smaller value may be chosen to reduce the computing time.
Then, an initial clustering configuration with $K_{up}$ groups is generated.
This may be created at random, or following initialisation methods based on the k-means algorithm, such as those described in \textcite{matias2017statistical} or \textcite{rastelli2017choosing}.
At this point the main routine of the algorithm starts, where an active node $\left( t,i \right)$ is selected, and its allocation is updated.
For the update, all possible moves to groups $1,\dots,K_{up}$ are tested, and, finally, the change yielding the best increase in the objective function is performed (note that the label zero remains exclusive of inactive nodes throughout).
This process continues in a loop until no further increase is possible.
After convergence, hierarchical clustering updates are attempted on the final solution obtained, following exactly the same procedure described in \textcite{come2015model} and \textcite{rastelli2017choosing}. 
The computing time demanded by this last step is usually negligible, yet it may improve the final solution by merging together some of the groups.
The pseudocode for the algorithm (called \texttt{GreedyICL}) is provided in Algorithm \ref{GreedyICL}. 
Note that, in the pseudocode, $\ell_{\left( t,i \right)\rightarrow \hat{g}}$ denotes the value corresponding to the current allocations with node $\left( t,i \right)$ moved to group $g$.
\begin{algorithm}[htb]
\begin{spacing}{1.2}
\caption{\texttt{GreedyICL}}
\label{GreedyICL}
\begin{algorithmic}
\State Set $K_{up}$ and initialise the allocations $\mathcal{Z}$.
\State Evaluate the objective function and set $\ell = \ell_{stop} = \log\left(\mathcal{ICL}_{ex}\right)$.
\State Set $stop = false$.
\While{$!stop$}
\State Set $\mathcal{U} = \left\{ \left( t,i\right):\ z_{i}^{(t)} \neq 0,\ t=1,\dots,T,\ i=1,\dots,N\right\}$.
\State Shuffle the elements of $\mathcal{U}$.
\While{$\mathcal{U}$ is not empty}
\State $\left( t,i \right) = pop\left( \mathcal{U} \right)$.
\State $\hat{g}=\arg \max_{\substack{g=1,2,\dots,K_{up}}} \ell_{\left( t,i \right)\rightarrow g}$.
\State $\ell=\ell_{\left( t,i \right)\rightarrow \hat{g}}$.
\State $z_i^{(t)} = \hat{g}$.
\EndWhile
\If{$\ell \leq \ell_{stop}$} $stop = true$ \textbf{else} $\ell_{stop} = \ell$.
\EndIf
\EndWhile
\State Return $\mathcal{Z}$ and $\ell$.
\end{algorithmic}
\end{spacing}
\end{algorithm}

Both \texttt{GreedyICL} and the final merge procedure only involve greedy updates, so they can only increase the objective function value. 
However, there is no guarantee that the final solution will correspond to a global optimum of $\log\left(\mathcal{ICL}_{ex}\right)$:
for this reason, several restarts of the whole procedure may be beneficial to avoid local optima.

From the algorithmic point of view, one main advantage of these greedy updates is their scalability: the increase in the objective function for each move can be evaluated very efficiently. 
Furthermore, convergence is usually reached after very few updates of each of the allocations.
More detailed explanations regarding the computational savings are provided for example in \textcite{come2015model} and \textcite{rastelli2017choosing} and references therein.

As already pointed out, the number of groups can be deduced from the allocation variables at any stage.
This makes \texttt{GreedyICL} particularly appealing, because, in one single algorithmic framework, one can obtain an estimate of the best $K$, according to the exact ICL criterion.
In fact, an advantage of the exact ICL approaches of \textcite{bertoletti2015choosing,come2015model,wyse2017inferring,rastelli2017choosing} 
is that they do not rely on a grid search over all possible $K$ values, which becomes impractical if the number of groups is large.

\section{Simulations}\label{sec:simulations}
In this section, artificial data is used to validate the methodology described in this paper.

\subsection{Simulation study 1}\label{sec:simulation_study_1}

\begin{figure}[htb]
\centering
\includegraphics[width=0.49\textwidth]{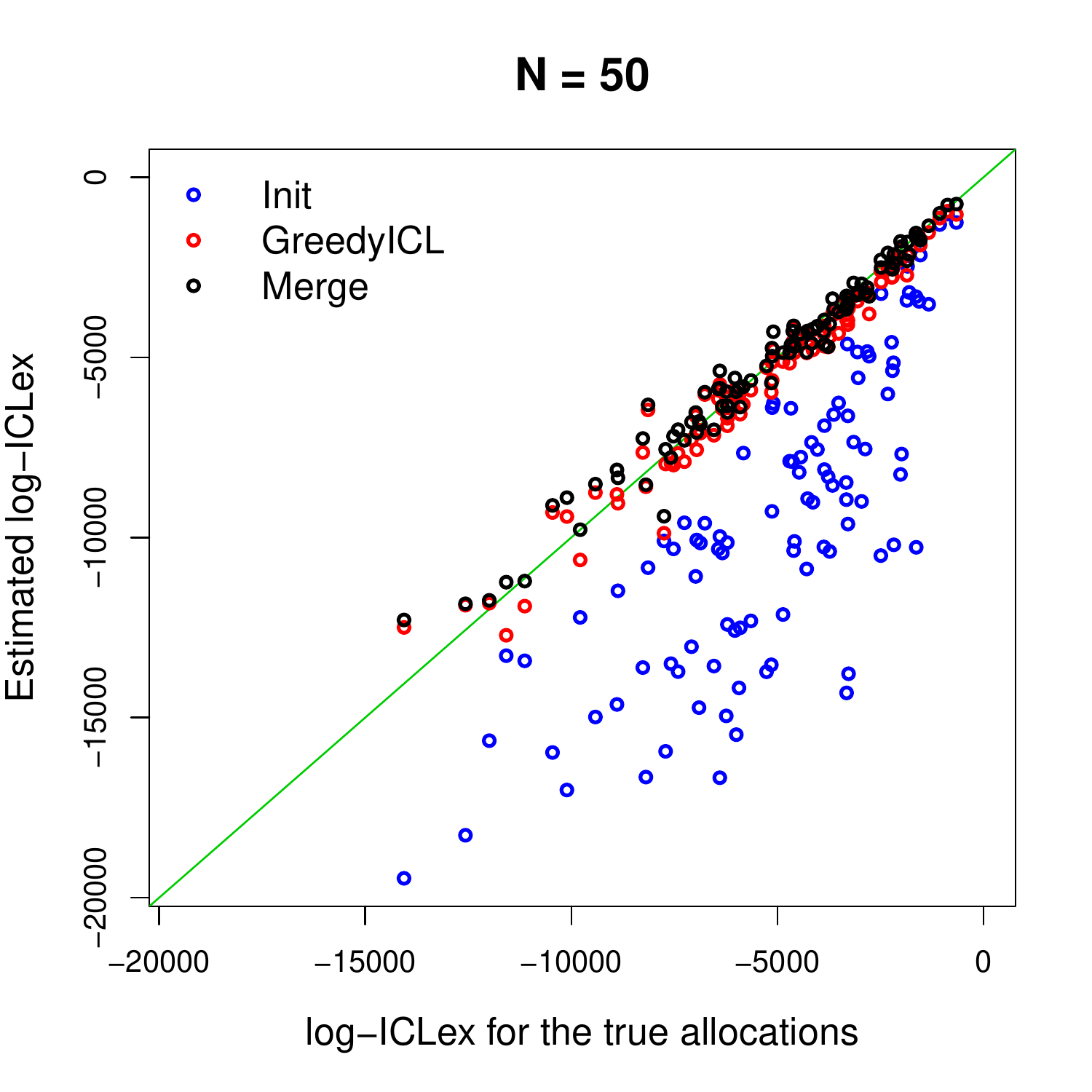}
\includegraphics[width=0.49\textwidth]{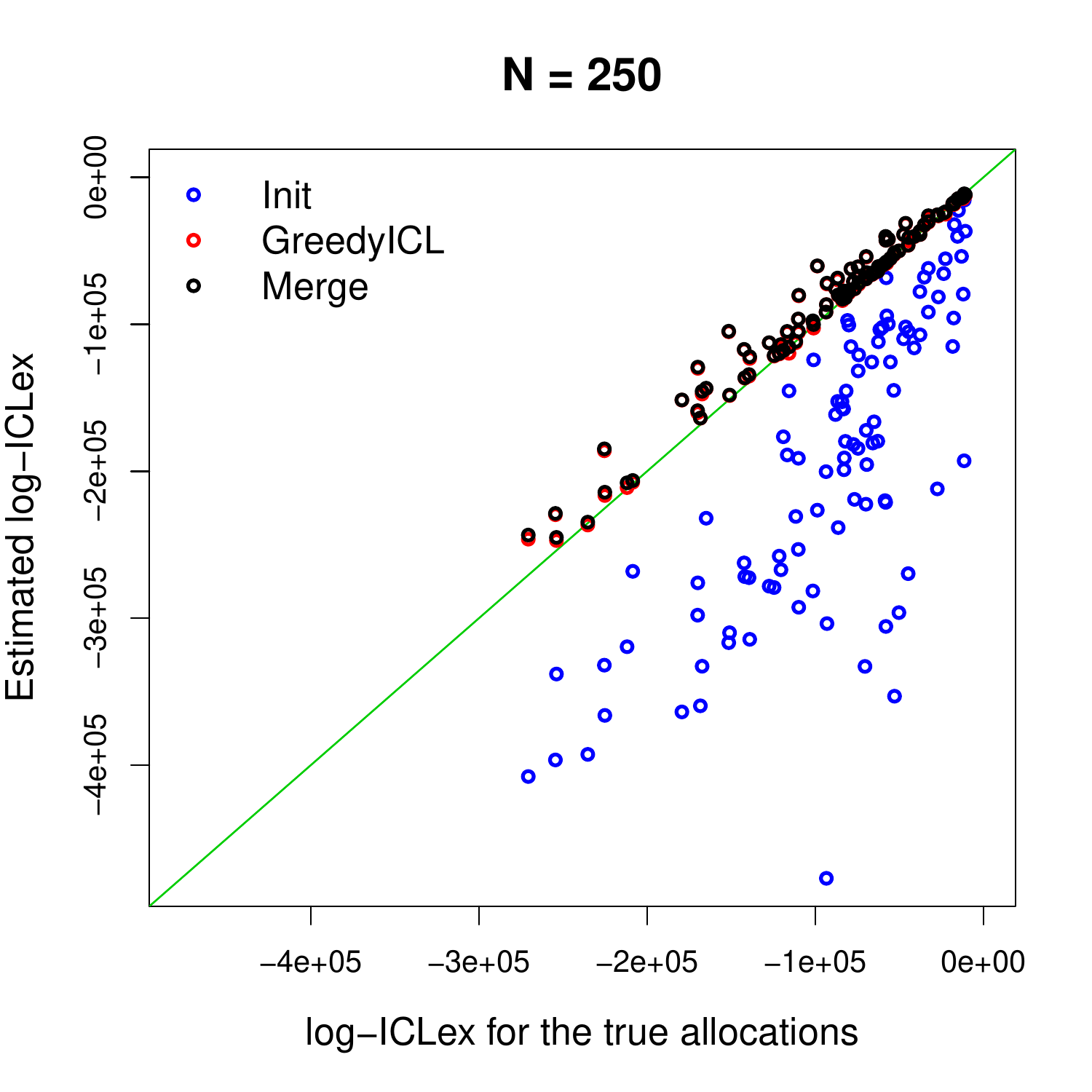}
\caption{\textbf{Simulation study 1}. $\log(\mathcal{ICL}_{ex})$ values for the true allocations (on the horizontal axes) and for the best estimated clustering across all $K_{up}$ values (on the vertical axes). 
The blue circles correspond to the values obtained after the initialisation using k-means, 
the red circles to those obtained after the \texttt{GreedyICL} described in \ref{GreedyICL} 
and the black circles correspond to the values obtained at the end of the merging procedure.}
 \label{fig:icl1}
\end{figure}

In the first simulated setting considered, the number of time frames is $T=20$, whereas two scenarios are possible for the number of nodes: $N=50$ or $N=250$.
The artificial networks are generated using the hierarchical structure described in Section \ref{sec:hierarchical}, with $K=3$ and the hyperparameters all set to $0.5$.
This means that the networks are generated through the same posterior distribution which is being optimised, with only $K$ being arbitrarily fixed.
$100$ networks are independently generated, and the methodology described in Section \ref{sec:greedy} is run on each of them, once for each $K_{up}$ in $\left\{ 10,20,30\right\}$.
Figure \ref{fig:icl1} shows the objective function values for the true allocations and for the estimated clustering after each of the steps of the optimisation.
For most datasets, and for both small and large networks, the final solution achieves better $\log(\mathcal{ICL}_{ex})$ values than the true clustering, suggesting good convergence.
Also, the increase granted by the \texttt{GreedyICL} step is generally much larger than that given by the final merge step.

\begin{figure}[htbp]
\centering
\includegraphics[width=0.49\textwidth]{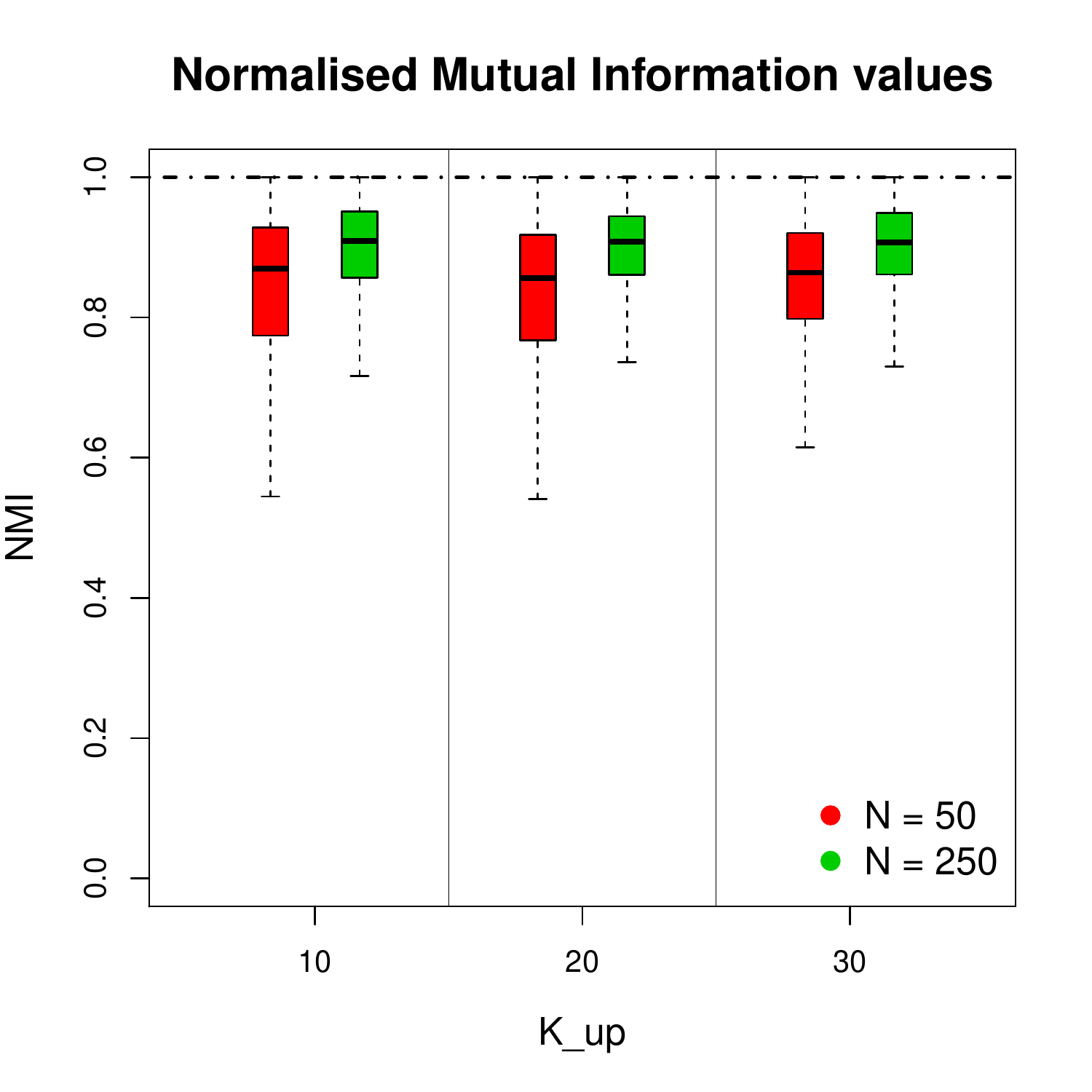}
\includegraphics[width=0.49\textwidth]{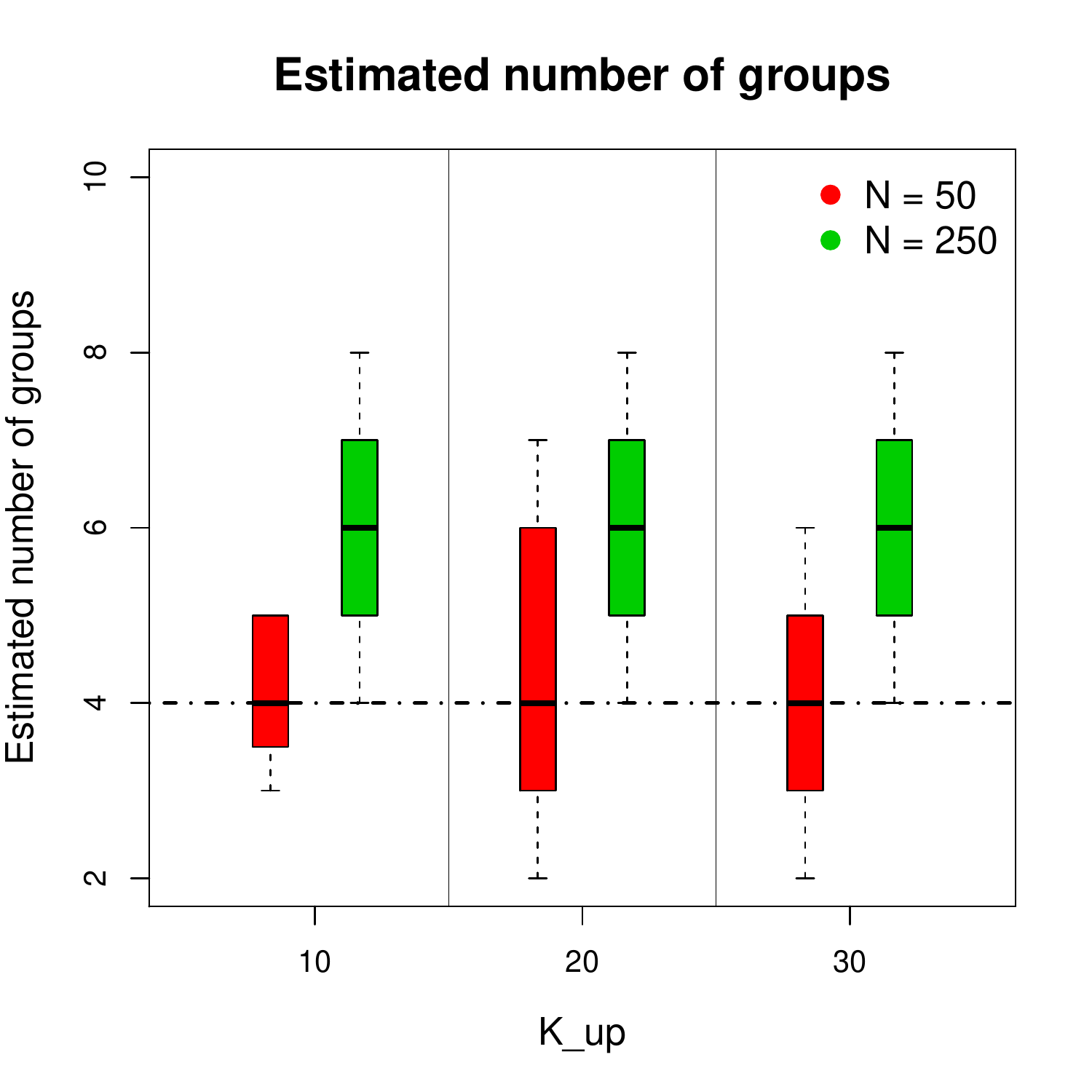}
\caption{\textbf{Simulation study 1}. The left panel shows the Normalised Mutual Information (NMI) index between the true clustering and the estimated clustering, for all combinations of $K_{up}$ and $N$.
The boxplots represent the values across all generated networks and time frames (the index is in fact evaluated for each $t=\left\{ 1,\dots,T\right\}$ independently).
The boxplots on the right panel are generated in the same way, and they show the estimated number of groups at each time frame.}
 \label{fig:k_nmi}
\end{figure}

Figure \ref{fig:k_nmi} focuses instead on the performance of the $\mathcal{ICL}_{ex}$ criterion. 
The Normalised Mutual Information Criterion \parencite{strehl2002cluster} is used to compare each estimated partition to its corresponding true counterpart.
The plot on the left panel of Figure \ref{fig:k_nmi} shows a very high level of agreement, particularly for the larger datasets.
Note that this criterion is normalised, hence it should not be affected by the value of $N$.

Regarding the optimal number of groups (shown on the right panel of Figure \ref{fig:k_nmi}), it seems that the criterion tends towards an overestimation, at least in larger networks. 
This may be related to the presence of more outliers which increase the heterogeneity of the data.
Finally, both plots highlight that the choice of $K_{up}$ does not affect performance. 
Note that a smaller $K_{up}$ reduces the computing time, yet the optimal partition can only be found if $K_{up}$ is greater than the optimal number of groups.
Hence, in general, the higher $K_{up}$ the better; nevertheless, smaller $K_{up}$ values may be used to speed up the algorithm or to force it to return a solution with fewer groups.

\subsection{Simulation study 2}\label{sec:simulation_study_2}
The second simulated setting aims at highlighting that the model proposed is fundamentally different from other available methods, such as that of \textcite{matias2017statistical}.
In fact, this section shows that the method proposed in this paper can achieve better performances in datasets that exhibit strong time dependencies and persistence of edges or non-edges.

In this simulated setting, the number of time frames is again set to $T=20$, whereas the number of nodes is set to $50$.
The three latent groups considered are characterised by the following edge probabilities:
\begin{equation}
\rowcolors{1}{}{}
 \boldsymbol{\Theta} = \left(\begin{matrix}
                        0.9 & 0.1 & 0.1\\
                        0.1 & 0.9 & 0.1\\
                        0.1 & 0.1 & 0.9\\
                       \end{matrix}\right)
                       ,\hspace{1cm}
 \textbf{P} = \left(\begin{matrix}
                        0.9 & 0.1 & 0.1\\
                        0.1 & 0.9 & 0.1\\
                        0.1 & 0.1 & 0.1\\
                       \end{matrix}\right)
                       ,\hspace{1cm}
 \textbf{Q} = \left(\begin{matrix}
                        0.1 & 0.1 & 0.1\\
                        0.1 & 0.9 & 0.1\\
                        0.1 & 0.1 & 0.9\\
                       \end{matrix}\right)
                       .
\end{equation}
The SBM-type probabilities simply follow a community structure with high within-groups probabilities. 
As concerns $\textbf{P}$ and $\textbf{Q}$, if two nodes belong to the same group, the situation can be summarised as follows: 
in group $1$ they tend to create edges frequently, but they seldom destroy them; 
in group $2$ they tend to create and destroy edges very frequently; 
whereas in group $3$ they destroy edges frequently but create them seldom.
Whenever the two nodes are in two different groups, they tend not to change the current state of their interaction.
Regarding the transition probabilities, the nodes remain in the same group with probability $0.8$ or can move to another group completely at random. 
However, the group of inactive nodes is inexistent in this case, in that nodes cannot ever become inactive.

Using this parameter configuration, $500$ networks are generated at random.
On each of these, the \texttt{GreedyICL} procedure is run once with $K_{up} = 10$, and the \texttt{dynsbm} procedure of \textcite{matias2017statistical} is run once for every choice of $K=1,\dots,6$.
While the \texttt{GreedyICL} method chooses the number of groups in one run, in \texttt{dynsbm} only the run corresponding to the highest approximate ICL is retained as optimal, as advised in the related paper.

\begin{figure}[htbp]
\centering
\includegraphics[width=0.49\textwidth]{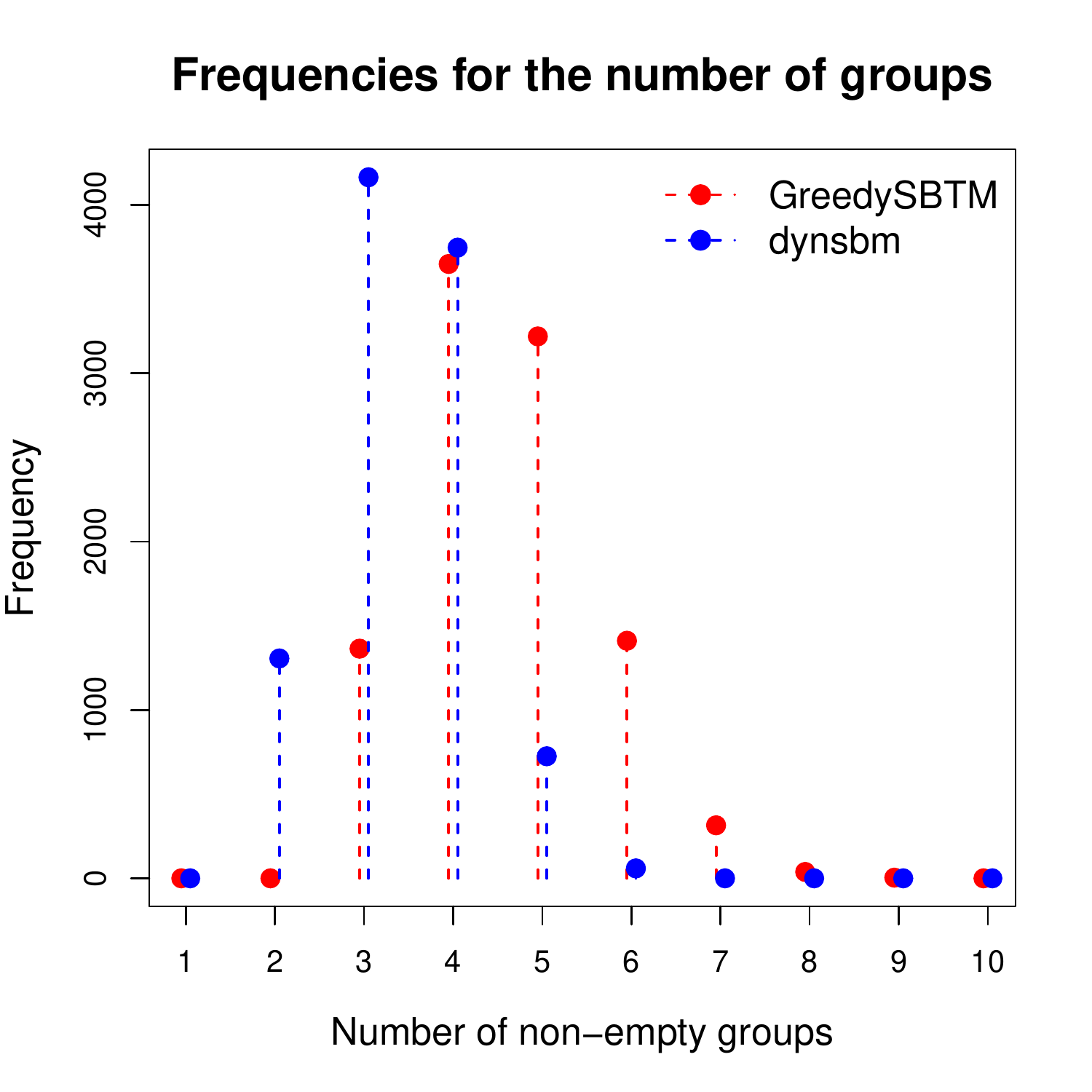}
\includegraphics[width=0.49\textwidth]{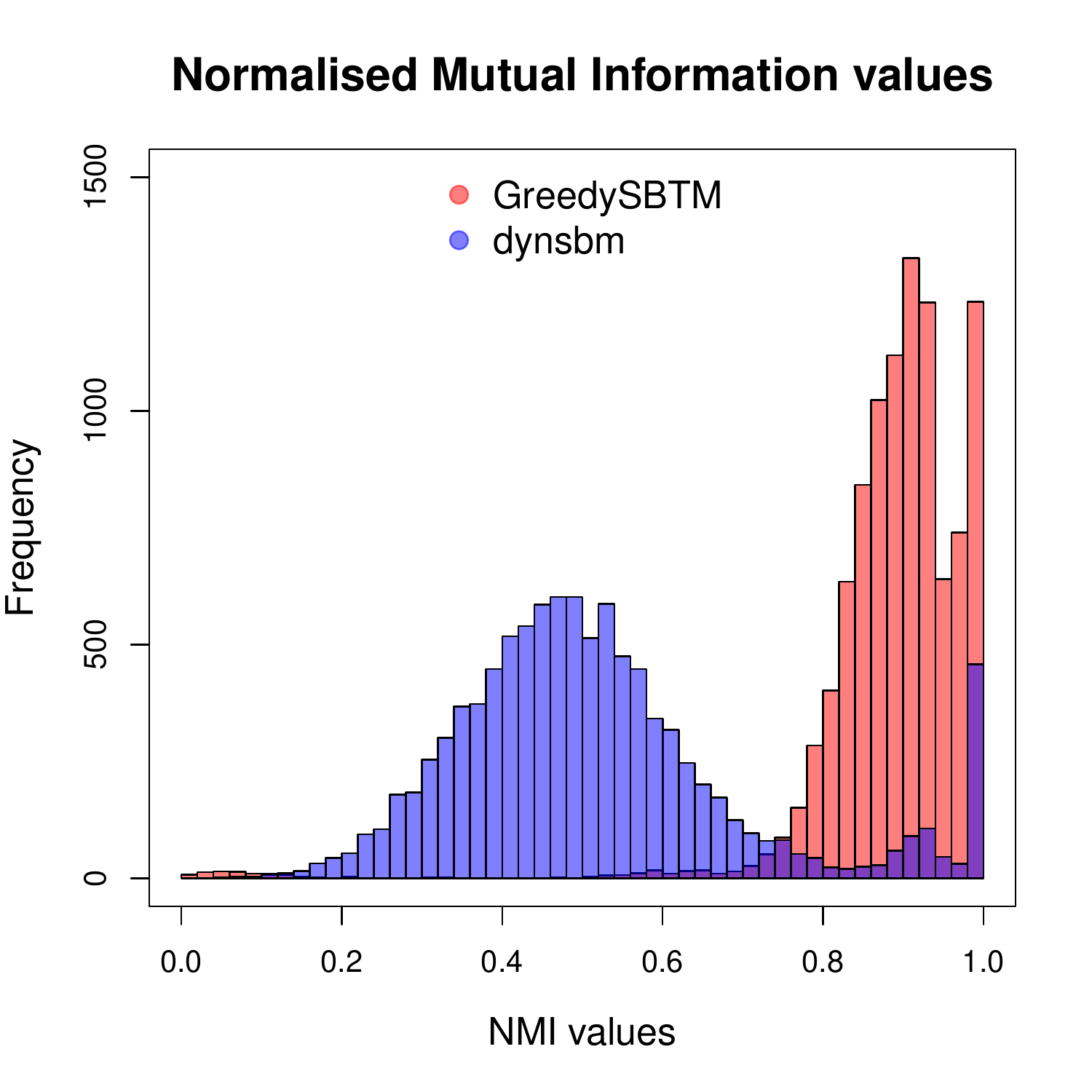}
\caption{\textbf{Simulation study 2}. The left panel shows the aggregated number of time frames in which the correct number of groups ($3$) was properly estimated, for the proposed greedy approach and for the algorithm \texttt{dynsbm} of \textcite{matias2017statistical}. 
On the right panel the Normalised Mutual Information (NMI) indexes between the true clusterings and the estimated clusterings are shown, for both methods and for each time frame independently.}
 \label{fig:sim_2}
\end{figure}

Figure \ref{fig:sim_2} illustrates the results obtained in this experiment.
Similarly to the previous simulation study, the \texttt{GreedyICL} tends towards an overestimation of the number of groups (see left panel of the figure), in that the correct value $K=3$ is properly estimated in about $15\%$ of the cases.
The \texttt{dynsbm} seems to achieve better performance in this task. 
However, as documented in the plot on the right panel, this event is rather fortuitous, since the vast majority of the optimal solutions of \texttt{dynsbm} are fundamentally wrong, and they do not capture the essence of the generative mechanism of the data.
In other words, the \texttt{dynsbm} method provides a different view on the data, which is not necessarily appropriate when high persistence of the edges is present.

Table \ref{tab:study_2} shows the average computing times for the two algorithms.
\begin{table}[htb]
\centering
\begin{tabular}{cc}
  \specialrule{.1em}{0em}{0em}
          \texttt{GreedySBTM} & 0.083  \\
          \texttt{dynsbm} & 14.959  \\
   \specialrule{.1em}{0em}{0em}
\end{tabular}
\caption{\textbf{Simulation study 2}. Seconds (rounded value) required to run once each of the algorithms.}
\label{tab:study_2}
\end{table}

\section{Reality Mining dataset}\label{sec:reality}
\begin{figure}[htb]
\centering
\includegraphics[width=0.49\textwidth]{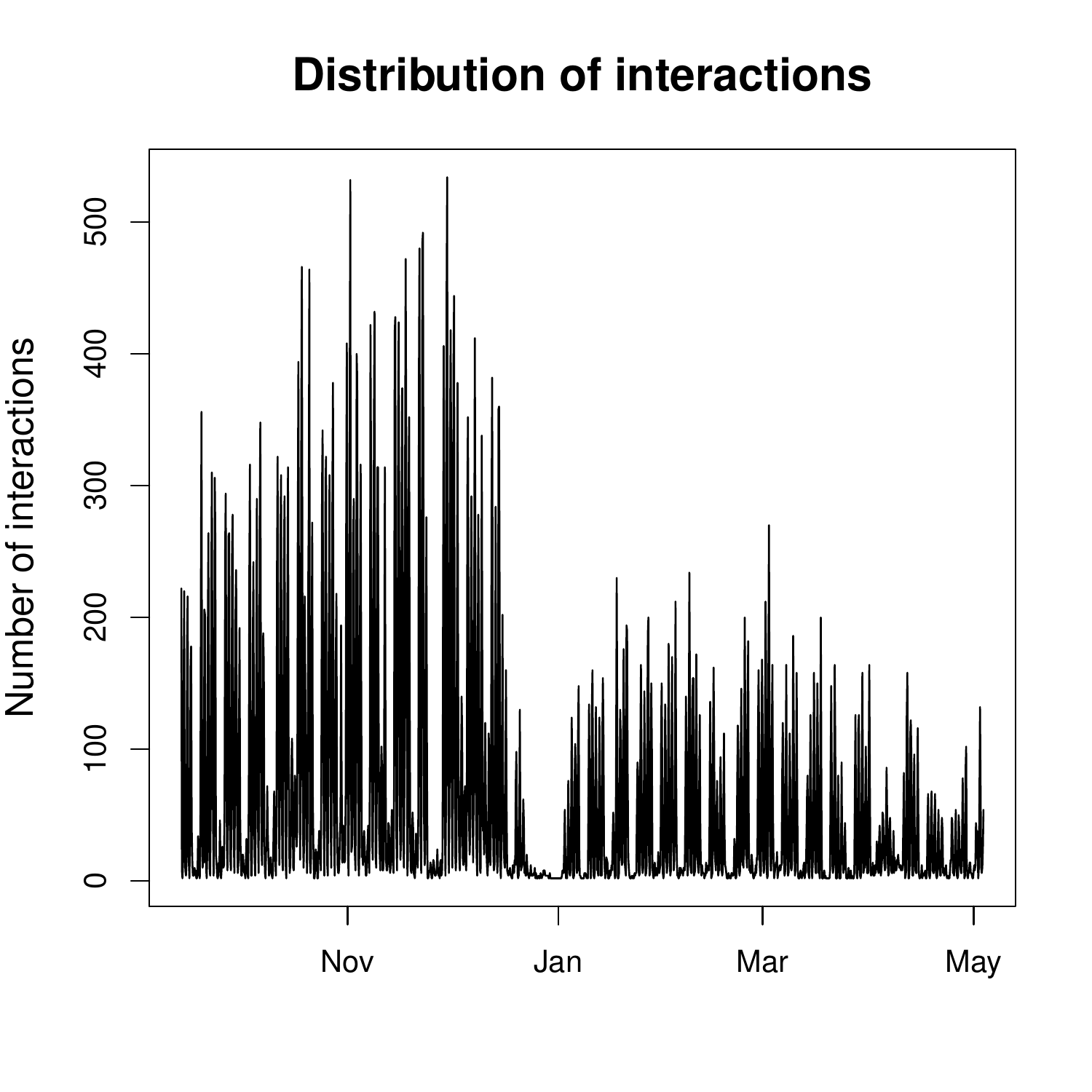}
\includegraphics[width=0.49\textwidth]{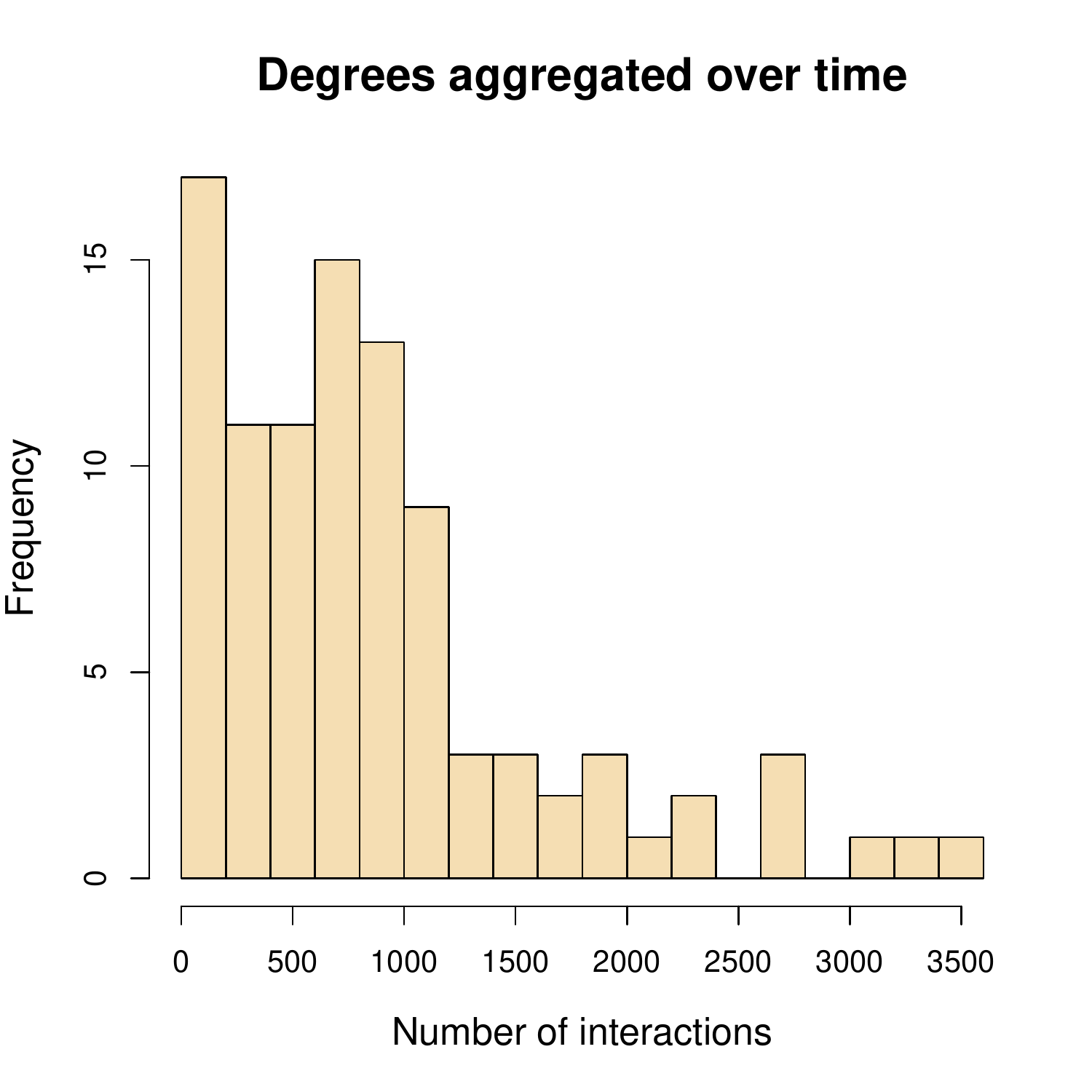}
\caption{\textbf{Reality mining dataset.} The plot on the left panel shows the number of edges at each of the time frames. The plot on the right panel shows instead the frequencies for the total number of edges incident to each node.}
 \label{fig:rm_times}
\end{figure}

The Reality Mining experiment was performed in $2004$ as part of the Reality Commons project.
The data was collected and first described by \textcite{eagle2006reality}, and it includes human contacts between Massachusetts Institute of Technology (MIT) students, from $14$ September $2004$ to $5$ May $2005$.
KONECT (the Koblenz Network Collection) provides a public version of a proximity network extracted from the Reality Mining data. 
The dataset describes proximity interactions of students through a list of undirected edges and their corresponding time stamp.
The number of nodes having at least one interaction is $N=96$, and the total number of interactions is $1{,}086{,}404$.
The $9$ months were discretised in $T=1392$ time frames of $4$ hours each.
Then, an adjacency cube $\mathcal{X}$ of size $N\times N\times T$ was created as follows:
\begin{equation}
\rowcolors{1}{}{}
 x_{ij}^{(t)} = \begin{cases}
                 1&\mbox{ if nodes $i$ and $j$ had at least one interaction between $t-1$ and $t$,}\\
                 0&\mbox{ otherwise. }
                \end{cases}
\end{equation}
The distribution of edges in the new representation is shown in Figure \ref{fig:rm_times}.
The nodes were considered inactive in all of the time frames where they had zero interactions: as a consequence, approximately $83\%$ of the allocation variables were set to zero, overall.
The algorithm was then run with $K_{up} = 20$ and it converged after $15$ iterations and $115$ seconds.

\begin{figure}[htb]
\centering
\includegraphics[width=0.49\textwidth]{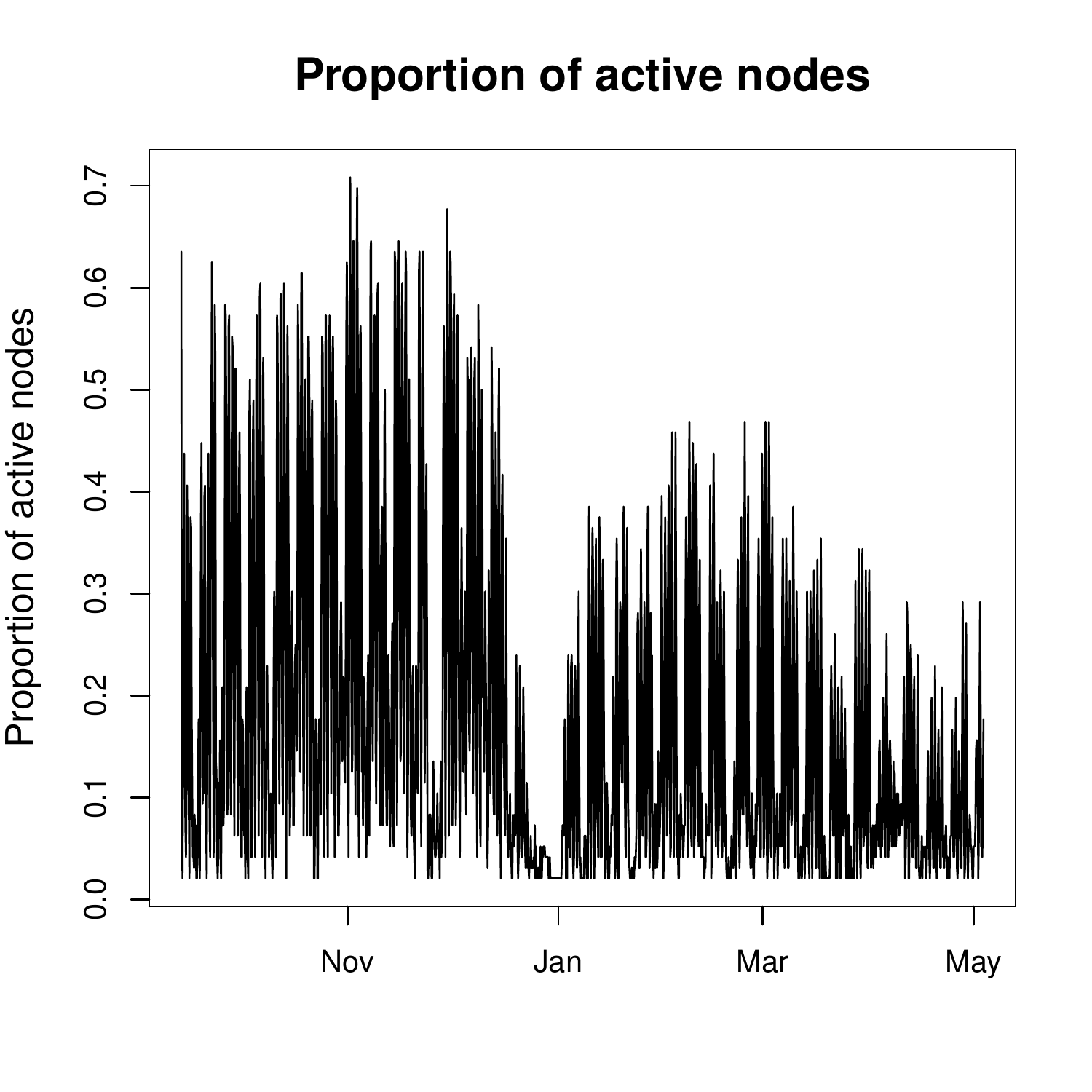}
\includegraphics[width=0.49\textwidth]{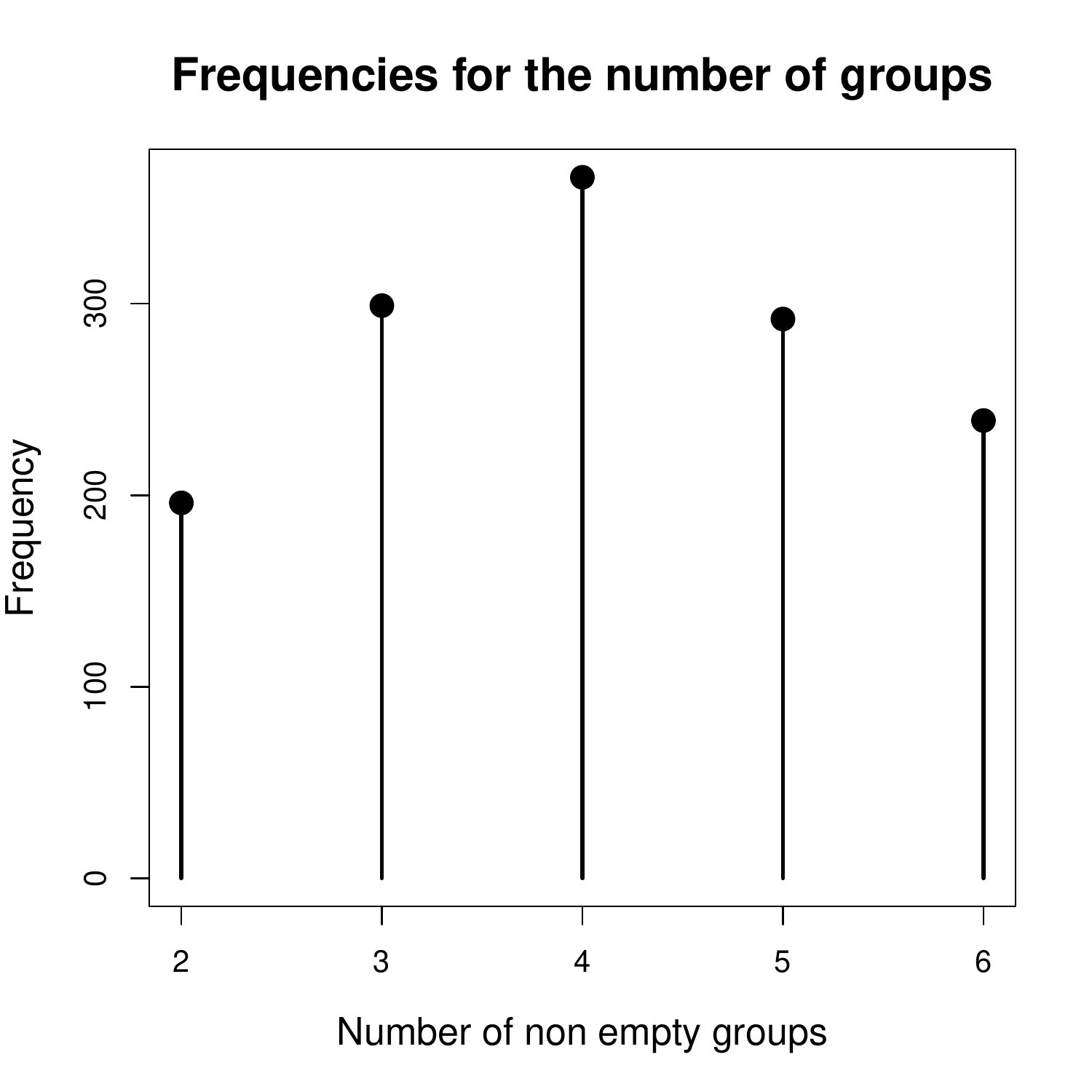}
\caption{\textbf{Reality mining dataset.} The plot on the left panel shows the proportion of nodes which are active in each time frame. 
The plot on the right panel shows instead the number of time frames where the number of non empty groups is equal to $k$, for $k=1,\dots,6$.}
 \label{fig:rm_k}
\end{figure}

\begin{figure}[htbp]
\centering
\includegraphics[width=0.45\textwidth, page=1]{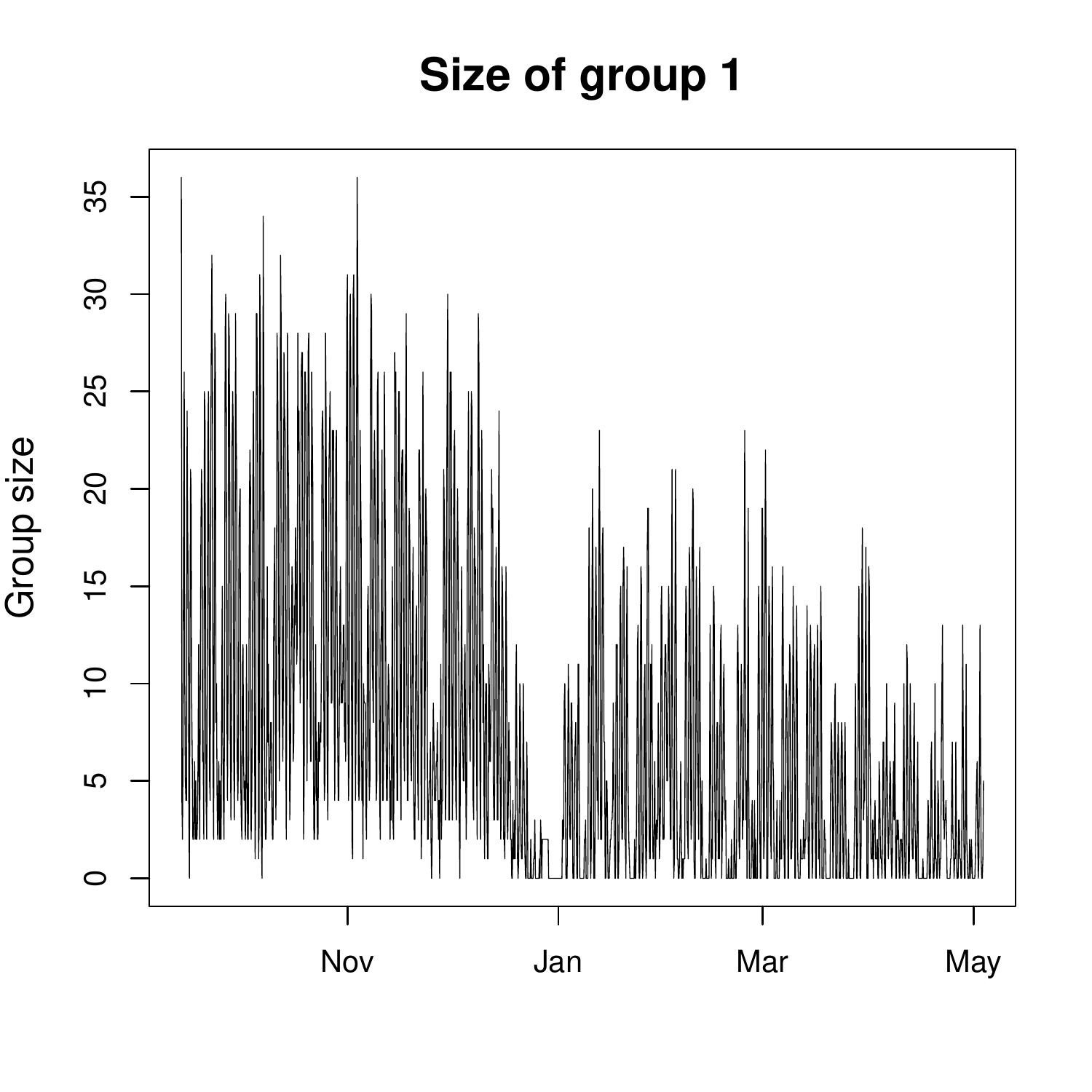}
\includegraphics[width=0.45\textwidth, page=2]{mit_counts.pdf}\\
\includegraphics[width=0.45\textwidth, page=3]{mit_counts.pdf}
\includegraphics[width=0.45\textwidth, page=4]{mit_counts.pdf}\\
\includegraphics[width=0.45\textwidth, page=5]{mit_counts.pdf}
\caption{\textbf{Reality mining dataset.} These plots show the number of nodes contained by each group at every time frame.}
 \label{fig:rm_counts}
\end{figure}

The resulting number of groups is $K = 5$, meaning that if a node is active, it will select one of $5$ different connectivity profiles at each time frame.
The sizes of the groups of active nodes aggregated over time are $N_{1} = 10{,}143$, $N_{2} = 5{,}940$, $N_{3} = 2{,}095$, $N_{4} = 4{,}315$, and $N_{5} = 547$.
Figure \ref{fig:rm_k} shows the frequencies of the number of groups across all of the time frames.
These plots suggest that very often several of the groups are empty, meaning that the network is temporarily homogeneous. 
This is emphasised in Figure \ref{fig:rm_counts}, where the size of all groups is shown for each time frame.
The migrations between groups exhibit a clear temporal pattern, mostly following the day/night cycle.
Additionally, a longer period of inactivity is observed at the end of December, where most nodes become inactive.

\begin{figure}[htb]
\centering
\caption*{$\hat{\textbf{P}} \hspace{5cm} \hat{\textbf{Q}} $}
\includegraphics[width=0.32\textwidth]{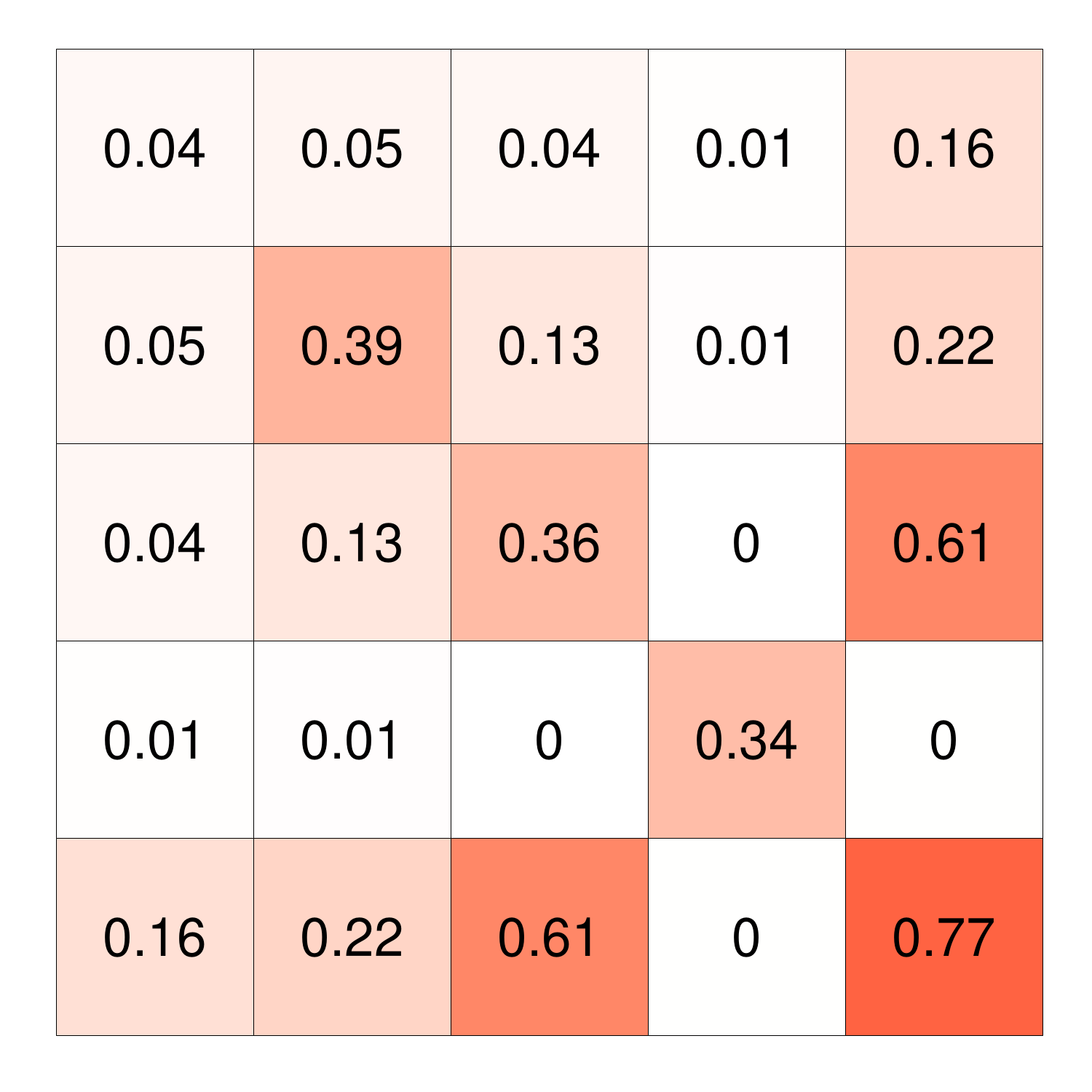}
\includegraphics[width=0.32\textwidth]{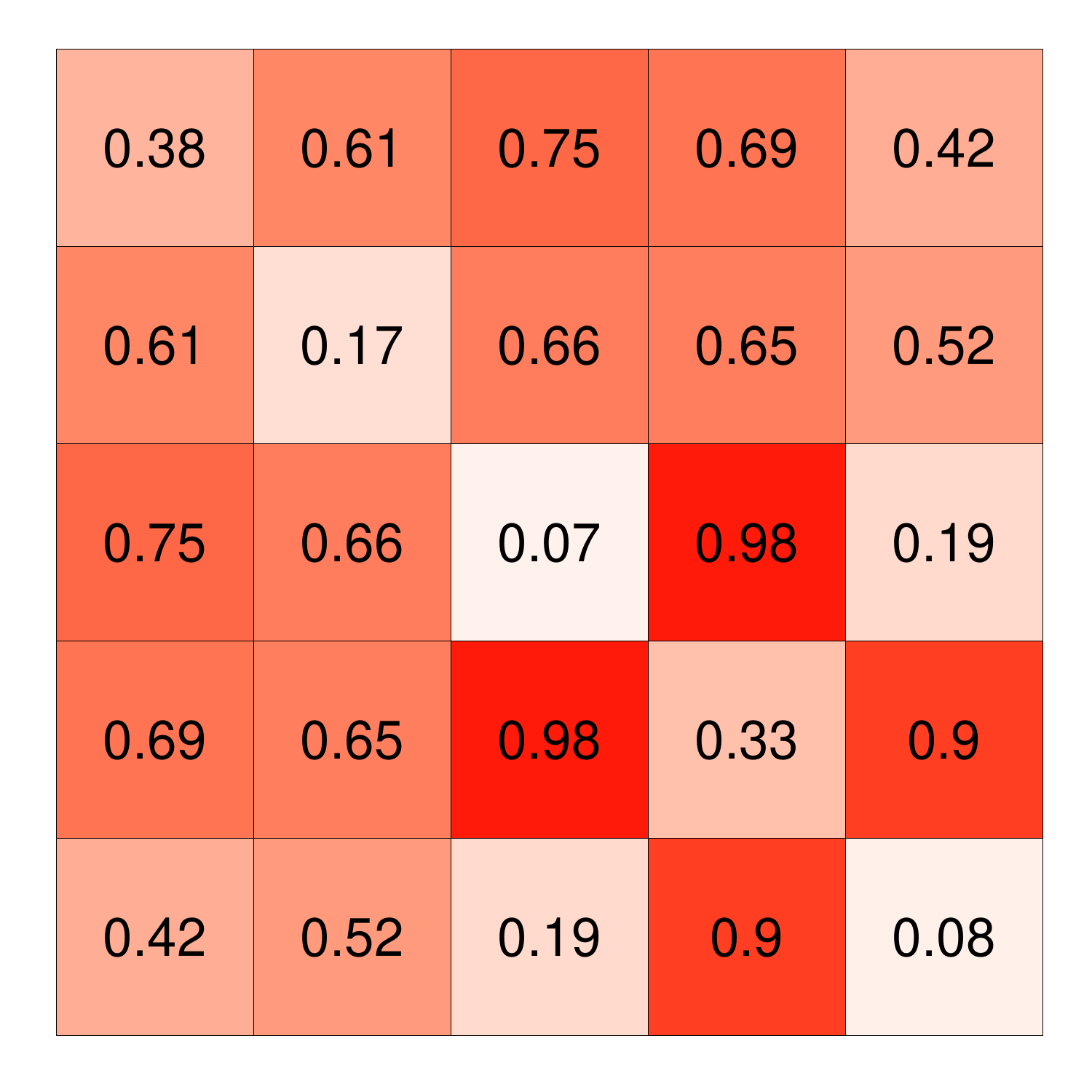}\\
\caption*{$\hat{\boldsymbol{\Theta}} \hspace{5cm} \hat{\boldsymbol{\Pi}} $}
\includegraphics[width=0.32\textwidth]{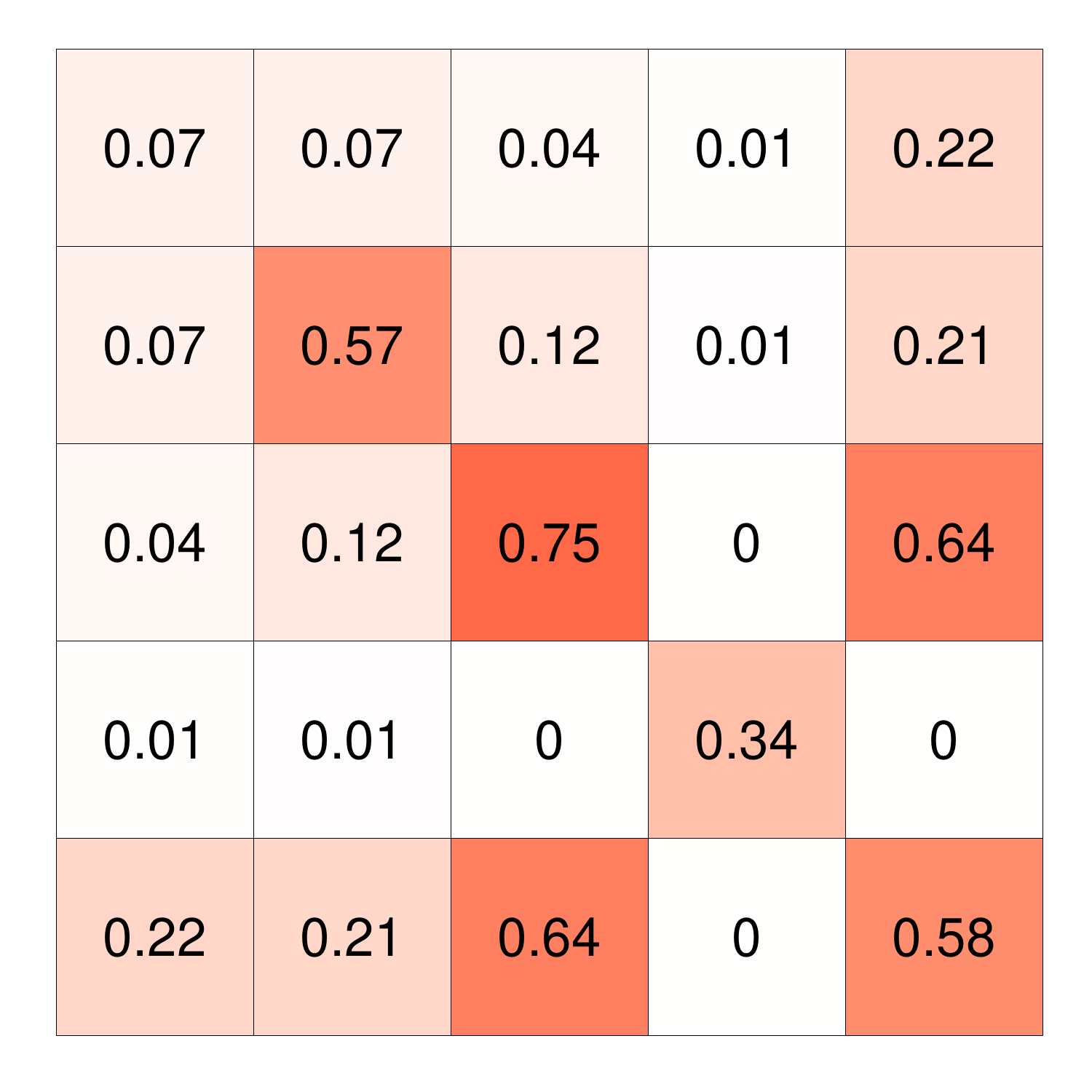}
\includegraphics[width=0.32\textwidth]{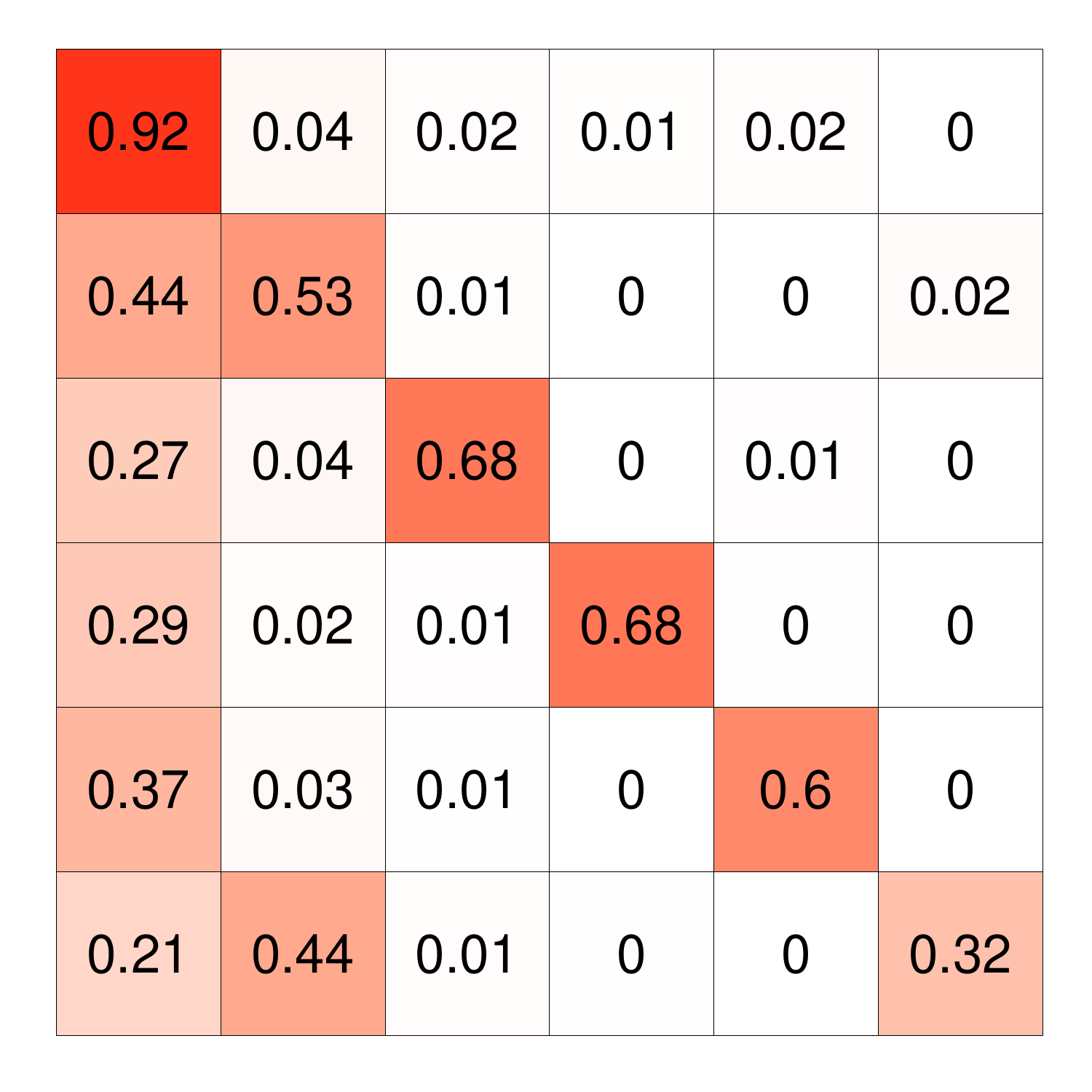}
\caption{\textbf{Reality mining dataset:} Estimated connection and transition probability matrices. The values for the group of inactive nodes are included only in $\hat{\boldsymbol{\Pi}}$, in the bottom-right corner.}
 \label{fig:rm_probs}
\end{figure}

Plug-in estimators for the connection probabilities are available as follows:
\begin{equation}
 \hat{P}_{gh} = \frac{U_{gh}^{01}}{U_{gh}^{01} + U_{gh}^{00}};\hspace{0.5cm}
 \hat{Q}_{gh} = \frac{U_{gh}^{10}}{U_{gh}^{10} + U_{gh}^{11}};\hspace{0.5cm}
 \hat{\theta}_{gh} = \frac{\eta_{gh}} {\eta_{gh} + \zeta_{gh} };\hspace{0.5cm}
 \hat{\pi}_{gh} = \frac{R_{gh}}{\sum_{h=0}^{K}R_{gh}}.
\end{equation}
For the Reality Mining dataset considered, these quantities are shown in Figure \ref{fig:rm_probs} through the matrices $\hat{\textbf{P}}$, $\hat{\textbf{Q}}$, $\hat{\boldsymbol{\Theta}}$ and $\hat{\boldsymbol{\Pi}}$, respectively.
Overall, the matrices $\hat{\boldsymbol{\Theta}}$ and $\hat{\textbf{P}}$ exhibit high values on the leading diagonal suggesting assortative behaviour and the presence of community structure.
The matrix $\hat{\textbf{Q}}$ exhibits the opposite situation, suggesting that edges are destroyed more frequently only if the nodes do not belong to the same group.
This is reasonable, since it implies that edges between nodes in the same group are created more frequently and kept for a longer time, confirming the presence of communities and edge-persistence.

One can combine the information from the these matrices to notice interesting disassortative patterns.
In group number one, for example, the diagonal element in $\hat{\textbf{P}}$ is small, and the nodes are more likely to connect with nodes in group five.
Group five is also particularly connected with groups two and three, suggesting that the nodes in this group act like hubs in the network.
By contrast, group four is the only one exhibiting a very strong community structure, since the nodes in this groups interact almost exclusively with each other. 
This group may correspond to a particular community which displays isolation from the rest of the network.

As concerns the transition probabilities, the matrix $\hat{\boldsymbol{\Pi}}$ exhibits high values on the diagonal which suggest high stability, since nodes tend not to change much their allocations over time.
This is particularly true for group zero, containing the inactive nodes, which also continuously attracts nodes from all other groups.

\section{Conclusions}\label{sec:conclusions}
This paper has introduced a new methodology to estimate the number of groups and the optimal clustering of the nodes in a Stochastic Block Transition Model.
The criterion optimised is the exact Integrated Completed Likelihood, which has recently also been adopted in several other network modelling contexts.
Such criterion is maximised using an iterative greedy procedure, which is known to be particularly computationally efficient.
One important advantage is that the method infers the number of latent groups within the same algorithmic framework, hence without requiring a grid search over all possible models.
Also, although the framework is Bayesian, a non-informative set of prior distributions may be used, therefore resembling a black-box procedure.

The generative process considered allows nodes to become temporarily or permanently inactive, making this approach appropriate for temporal networks with very many time frames.
Crucially, the inactivity of the nodes is modeled in a very natural way, which can potentially lighten the computational burden substantially.

The procedure has been applied to both artificial and real dataset, showing that it can scale well with the size of the data.
The simulation studies have shown that the method usually converges to excellent solutions, yet in larger datasets it may overestimate the number of groups.
This seems to be a weak spot for the exact ICL criterion itself, since similar issues may be argued in other related works, such as \textcite{rastelli2017choosing}.
In addition, the simulations highlight that other available methods that do not account for edge persistence may fail to capture the true generative mechanisms of the data,
and hence lead to qualitatively different clustering solutions and interpretations.

The application to the Reality Mining dataset offers a demonstration of the results that can be obtained. 
In this dataset, intense interaction periods appear to be distinctly fragmented due to recurring intervals of inactivity. 
The modelling approach proposed in this paper can handle this scenario in a natural way, and, more importantly, it can exploit the presence of inactive nodes to mitigate the computational burden.

This paper has focused on undirected binary dynamic networks only. 
It may be interesting to extend this approach to the non-binary case, and to find a way to model the transitions on the edge values that allow for computationally efficient inferential procedures.
Also, the discretisation of the time dimension may have a non-negligible effect on the data analysis: this is for example highlighted in \textcite{matias2015semiparametric}. 
Hence, another important future step would be to extend the Stochastic Block Transition Model principle to networks that evolve continuously on time.

Regarding the inferential procedure, several alternatives may be considered: 
similarly to \textcite{matias2017statistical}, a variational Expectation-Maximisation algorithm may be employed to find the latent clustering and the model parameters within the same algorithmic framework;
or, following the approaches of \textcite{wyse2012block,mcdaid2013improved,white2016bayesian}, a collapsed Gibbs sampler may be used to sample the allocations from their marginal posterior distribution, 
hence obtaining an assessment of the uncertainty regarding both clustering and number of groups.

The initialisation of the algorithm remains a very central issue, since the procedure is known to be sensitive to initial conditions, and the final solutions may potentially differ a lot between various restarts.
This paper uses the same initialisation method of \textcite{matias2017statistical} and \textcite{rastelli2017choosing}, however other possibilities (such as spectral clustering) may be explored.

The \texttt{R} package \texttt{GreedySBTM} accompanies this paper: 
it contains a \texttt{C++} implementation of the algorithms described in Section \ref{sec:greedy}, and it includes the adapted version of the Reality Mining dataset used in Section \ref{sec:reality}.
The package is publicly available on CRAN \parencite{rcoreteam}.

\section*{Acknowledgements}
The author thanks Kevin Xu for the useful conversations.
This research was supported through the Vienna Science and Technology Fund (WWTF) Project MA14-031.

\printbibliography

\end{document}